%% file: github.tex
\documentclass[10pt, conference]{IEEEtran}
\IEEEoverridecommandlockouts
\usepackage{cite}
\usepackage{amsmath,amssymb,amsfonts}
\usepackage{algorithmic}
\usepackage{graphicx}
\usepackage{textcomp}
\usepackage{xcolor}
\def\BibTeX{{\rm B\kern-.05em{\sc i\kern-.025em b}\kern-.08em
    T\kern-.1667em\lower.7ex\hbox{E}\kern-.125emX}}

\usepackage[export]{adjustbox}
\newcommand{\emoji}[1]{\includegraphics[width=1em,valign=t]{emoji_images/#1.png}}

\usepackage{subfigure}
\usepackage{multirow}
\usepackage{balance} 
\usepackage{url}
\begin{document}

%\title{Understanding Emoji Usage on GitHub: An Empirical Study\\
%%{\footnotesize \textsuperscript{*}Note: Sub-titles are not captured in Xplore and should not be used}
%%\thanks{Identify applicable funding agency here. If none, delete this.}
%}
\title{A First Look at Emoji Usage on GitHub: An Empirical Study}

%\author{\IEEEauthorblockN{Hidden for Double-blind Review}}
\author{\IEEEauthorblockN{Xuan~Lu~~ Yanbin~Cao~~ Zhenpeng~Chen~~ Xuanzhe~Liu~}
\IEEEauthorblockA{\textit{Key Laboratory of High Confidence Software Technologies (Peking University), Ministry of Education, PRC} \\
\textit{\{luxuan, caoyanbin, czp, xzl\}@pku.edu.cn}}}
%\and
%\and
%\and
%\and
%\and
%\author{
%Xuan~Lu~ Yanbin~Cao~ Zhenpeng~Chen~ Xuanzhe~Liu\\
% \affaddr{Key Laboratory of High Confidence Software Technologies (Peking University), Ministry of Education, PRC}\\
%      \email  \{luxuan, caoyanbin, czp, xzl\}@pku.edu.cn
%      }
\maketitle

\input{sections/abstract}

%\begin{abstract}
%This document is a model and instructions for \LaTeX.
%This and the IEEEtran.cls file define the components of your paper [title, text, heads, etc.]. *CRITICAL: Do Not Use Symbols, Special Characters, Footnotes, 
%or Math in Paper Title or Abstract.
%\end{abstract}

\begin{IEEEkeywords}
emoji, GitHub, developer, sentiment
\end{IEEEkeywords}

\input{sections/introduction}
\input{sections/related}
\input{sections/dataset}
\input{sections/characteristics}
\input{sections/interpretation}
%\input{sections/rq0_popularity}
%\input{sections/popularity}
%\input{sections/rq1_interpretation}
%\input{sections/rq2_characteristics}
%\input{sections/descriptive}
\input{sections/purpose}
\input{sections/implication}

\input{sections/threats}

\input{sections/conclusion}
%\input{sections/ack}
\balance{}

\bibliographystyle{IEEEtran}
\bibliography{github}

\end{document}

%% file: sections/abstract.tex
\begin{abstract}
Emoji is becoming a ubiquitous language and gaining worldwide popularity in recent years including the field of software engineering (SE). As nonverbal cues, emojis are widely used in user understanding tasks such as sentiment analysis, but few work has been done to study emojis in SE scenarios. This paper presents a large scale empirical study on how GitHub users use emojis in development-related communications. We find that emojis are used by a considerable proportion of GitHub users. In comparison to Internet users, developers show interesting usage characteristics and have their own interpretation of the meanings of emojis. In addition, the usage of emojis reflects a positive and supportive culture of this community. Through a manual annotation task, we find that sentimental usage is a main intention of using emojis in issues, pull requests, and comments, while emojis are mainly used to emphasize important contents in README. These findings not only deepen our understanding about the culture of SE communities, but also provide implications on how to facilitate SE tasks with emojis such as sentiment analysis.
\end{abstract}

%% file: sections/introduction.tex
\section{Introduction}

Emoji, defined as ``\textit{a digital image that is added to a message in electronic communication in order to express a particular idea or feeling}'',\footnote{\url{https://dictionary.cambridge.org/us/dictionary/english/emoji}} is emerging as a ubiquitous language with its compact visual and live presentation, rich semantics, and understandability, and gaining worldwide popularity in recent years. As nonverbal cues, emojis have been widely adopted on online communication services such as Twitter and Facebook, and various efforts have been made over emojis, in terms of  sentiment analysis~\cite{Vidal:2016fl, chen2018twitter}, user profiling~\cite{chen18}, personality assessment~\cite{marengo2017assessing}, and even cluture difference analysis~\cite{Lu:UbiComp16}. 

%However, these studies are often conducted with common Internet users on platforms like Twitter, without penetrating into a specific domain such as software engineering (SE).

A noticeable trend is that, emojis are not only widely adopted in regular online communication, but also increasingly used in software development practice, ranging from code in programs,\footnote{\url{https://www.tjvantoll.com/2016/06/10/emoji-and-coding/}}\footnote{\url{http://www.emojicode.org/} } to online forums such as StackOverflow and GitHub, and even to the requirements engineering~\cite{siadati2017modelling}. Given that emojis are expected to provide a new way to enrich the expression, increase the interaction vitality, and improve the communication efficiency, %and enrich the sentiment representation, 
we are motivated to understand how and why emojis are used in software engineering practice.
%from two aspects: (1) how and why emojis are used by software developers, and (2) whether emojis can help the communication and collaboration in software projects. 

%Since the use of emoji can reflect cultural difference among certain groups of users~\cite{Lu:UbiComp16}, can emojis be used to understand developers and the culture of the software community?

 %Given that sentiment analysis in SE tasks are limited to current tools~\cite{Novielli2015The, lin2018sentiment}, can emojis alleviate this problem? 

 %Despite this opportunity provided by emojis, very few studies have made done to understand the usage of emojis in software engineering fields and compare with the common Internet communities.

 %After all, only when the emojis reflect domain specific characteristics can they contribute to addressing domain specific problems.

In this paper, we conduct an empirical study of emoji usage in the software engineering community by exploring GitHub,\footnote{\url{https://github.com}} the most popular online community where millions of programmers host and manage projects and build software collaboratively, to study the way that they communicate with one another in development activities. More specific, we aim to answer the following research questions.  

\noindent \textbf{RQ1: How are emojis used by developers on GitHub?} We begin with investigating the characteristics of emoji usage on GitHub. We establish a large-scale data set of communicational posts (i.e., \textit{issue}, \textit{issue comments}, \textit{pull requests}, \textit{pull request comments}, and \textit{README}) collected from GitHub, spanning 66 months since January 2012. We make the descriptive analysis of the favored emojis, the density and position of emojis in GitHub posts, and derive some typical patterns. Domain specific usage can be observed, e.g., emojis can be assembled to represent \textit{slangs} such as ``\emoji{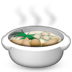}\emoji{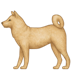}'' (i.e., \textit{dogfood}).

\noindent \textbf{RQ2: Do emojis have domain-specific meanings by developers on GitHub?} Suppose that the topics on GitHub can be quite different from daily communication due to the technical nature of the platform, we are interested in there are unique usage specific to the software development activities. We explore the meaning and sentiment of words that can be domain specific~\cite{Islam:2018hz}. Regarding that emojis are widely used as complements or surrogates of plain texts, we further leverage the state-of-the-art word embedding method for emoji interpretation in the technical context. Interestingly, the semantics of emojis on GitHub can be quite different from that on Twitter. For example, emojis such as \emoji{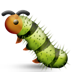} and \emoji{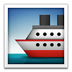} can present domain specific meanings. Results also indicate the tendency to express positive sentiments through emojis on GitHub.  %Interestingly, the emoji usage of GitHub users is significantly different from that of users on Twitter, indicating the tendency to express positive sentiments through emojis. For example, emojis such as \emoji{469} and \emoji{944} present domain specific meanings. Usage patterns of emojis in density, position, and selection in different posts are derived, indicating that emojis can be assembled with domain-specific meanings such as ``\emoji{308}\emoji{463}'' (i.e., \textit{dogfood}).

\noindent \textbf{RQ3: What are the intentions of using emojis by developers on GitHub?} %In common online communications, emojis are mostly used to express sentiments in daily communication~\cite{Hu2017Spice}, possibly due to the sentimental origins of emoji designs. 
Given the increasing importance of sentiment analysis in software development and the lack of efficient tools in this field~\cite{Novielli2015The, lin2018sentiment}, we expect that emojis can provide a new complementary signal to understand developers' sentiment in practice. We conduct an annotation task and build an eight dimension taxonomy of emoji usage intentions. We find that sentimental usage is a main intention in issues, pull requests, and comments, while emojis are mainly used to emphasize important content in README. In addition, sentimental emojis can also be used for non-sentimental intentions, indicating the significance of intention recognition before further application of emojis in analysis tasks in the SE field.

 %In particular, we provide detailed examples as demonstrations. 

%\textbf{Contributions}. Within our knowledge, we are the first to conduct an emoji usage analysis in the field of software engineering using a large-scale GitHub data set. We provide a comprehensive empirical analysis of the meaning, the usage pattern, and the intentions of emojis in developers' communications. We leverage the state-of-the-art word embedding method for emoji interpretation in the technical context, and release the embedding results to facilitate further research. We conduct a manual annotation task on 2,000 posts, propose the first taxonomy of emoji usage intention for GitHub, and display the distribution of intentions. Implications are provided for SE researchers as well as developers and platform operators. 

\noindent \textbf{Findings}. By answering the preceding questions, we find that developers use a diversity of emojis in their communications with domain-specific meanings and usage patterns. The use of emojis on Github reflects a positive and supportive culture of this community. Sentimental usage is still a main intention of using emojis on GitHub.

\noindent \textbf{Contributions}. To the best of our knowledge, this paper makes the first step of exploring emoji usage on software engineering activities and further understanding the behavior of the software developers with a new signal. Our derived findings and implications can probably help various stakeholders on how to better make use of emojis on GitHub, so as to improve the representation vitality, the communication efficiency, the possible sentiment inference, and to promote better project collaboration, problem solving, and productivity.

%To answer the aforementioned research questions,  We use the state-of-the-art word embedding methods to analyze the semantics and sentiments of emojis, so that comparisons between GitHub and the non-technical community can be made. We conduct a manual annotation task on 2,000 posts, so that a reliable taxonomy of emoji usage intention can be proposed.

%\textbf{Findings}. With the large-scale data set, we find that the preference for emojis of GitHub users is significantly different from Twitter, indicating the tendency to express positive sentiments through emojis. Various emojis such as \emoji{469} and \emoji{944} present domain specific meanings. Usage patterns of emojis in density, position, and selection in different posts are derived, including that emojis can be assembled with domain specific knowledge such as ``\emoji{308}\emoji{463}'' (i.e., \textit{dogfood}). The intention taxonomy based on annotation is finally constituted with eight dimensions, among which the sentiment effectiveness of emojis is most heavily relied on in conversational posts, while in README emojis are mainly used in content emphasis. In addition, emojis are widely used to smooth communications, which can be of significance to a positive and friendly collaboration platform. 

The rest of the paper is organized as follows. Section~\ref{sec:related} introduces related work. Section~\ref{sec:data} describes the data used in this study and demonstrates the popularity and distribution of emojis in our data set. Section~\ref{sec:characteristics} characterizes emoji usage patterns on GitHub. Section~\ref{sec:interpretation} develops an emoji interpretation method based on state-of-the-art embedding technique. %, indicating domain specific usage of emojis. 
%Section~\ref{sec:pattern} presents emoji usage patterns from the density, position, and selection of emojis in communications. %To further understand the domain specific meanings and usage patterns, 
Section~\ref{sec:intention} proposes a taxonomy of intentions of using emojis with a manual annotation task. Section~\ref{sec:implication} provides implications. Section~\ref{sec:threats} discusses threats to validity and future direction. Section~\ref{sec:conclusion} concludes the paper.

%% file: sections/related.tex
\section{Related Work}\label{sec:related}
%Our research originated from asking how could we attract more people to the developer community and is particularly related to three streams of existing literature: emoji usage analysis and participation in open-source communities.

%Yet the stereotype of program developer community place a bar between the public. Such stereotype hurts the broader participation in and impedes the diversity of the community~\cite{Doube:2012gz,cheryan2009ambient}.

%Our research is particularly related to three streams of existing literature: emoji usage analysis, language style in online communities, and participation in open-source communities. 

\subsection{Emoji Usage Analysis}
Increasingly popular, emojis are becoming a ubiquitous language and widely adopted by Internet users in recent years. Various studies have been done to analyze emoji usage across countries~\cite{LjubesicF16}, across cultures~\cite{Lu:UbiComp16}, and across demographic groups~\cite{chen18}. These research on emoji usage are mainly conducted in input methods~\cite{Lu:UbiComp16,Ai:2017wx,chen18}, instant messaging apps such as WhatsApp~\cite{al2015forms} and WeChat~\cite{zhou2017goodbye}, and social networks~\cite{barbieri2016does, Novak:2015}. So far, there is no study to analyze emoji usage and effects in a tech community. In addition, the prevalence of emojis also attracted many researchers to study the intentions and effects of using them. These research demonstrated that besides replacing the content words in text, emojis can also be used to provide emotional or situational information, adjust tones, express irony, engage the audience, decorate texts, etc~\cite{Hu2017Spice,Cramer:2016,PohlDR17}. However, due to the lack of emoji studies in tech community, we don't know which role emojis play and what functions emojis have in it.  To bridge the knowledge gap, we make the first effort to measure the usage of emojis on Github.

\subsection{Sentiment Analysis in Software Engineering} %Social Programmer Ecosystem
Understanding sentiments of developers is a research focus in software communities~\cite{sinha2016analyzing, robinson2016developer, pletea2014security, lin2018sentiment, JongelingDS15, Islam:2018hz,AhmedBIR17, DBLP:conf/esem/IslamZ17, DBLP:conf/msr/GuzmanAL14}. The sentiments of developers can be a sensor of the contributors' status and activity, as well as the quality of projects. %are related with activities of both the developer and the project. 
For example, sentiment analysis can be used to detect the psychological state and job satisfaction of developers~\cite{rousinopoulos2014sentiment}, which are strongly associated with their productivity and task completion quality. %On the other hand, sentiment analysis of developer-written messages and users' reviews can benefit the quality analysis of different projects and products, which is important for the downstream applications such as the recommendation of software libraries. 
However, current sentiment analysis tools have been demonstrated to have strong limitations on sentiment analysis in the SE field~\cite{lin2018sentiment}. In fact, to improve the sentiment analysis technique, many researchers in natural language processing community have started to use emojis as weak sentiment labels~\cite{ZhaoDWX12, DBLFelboMSRL17}, which is called distant-supervised learning. Such practice with emojis sheds lights on the improvement of SE sentiment analysis tools. To discover the possibility of leveraging emojis in sentiment analysis in SE, we first describe how emojis are used with the data set collected from GitHub in this work. 

%% file: sections/dataset.tex
\section{Data Set}\label{sec:data}
To investigate emoji usage in communications on GitHub, we select five typical types of posts, i.e., issues, issue comments, pull requests, pull request comments,\footnote{In general, an issue opens up a discussion thread for bugs, enhancement, questions, etc., while a pull request is a request for the project owner to ``pull'' the source code change from contributors and merge to the code repository. See https://help.github.com/categories/collaborating-with-issues-and-pull-requests/, retrieved in August 2018.} and README.\footnote{README helps developers to communicate expectations for their projects, see https://help.github.com/articles/about-readmes/, retrieved in August 2018.} To simplify, we refer to the issues, pull requests and their comments as \textit{conversational posts} for they can be replied by users. 

\subsection{Data Collection}
Through the GHTorrent project~\cite{Gousi13}, we collect the conversational posts spanning 66 months from January, 2012 to June, 2017. After a data cleaning process to filter duplicates, spams, and expired URLs, the data set covers 3,088,360 projects and 3,952,924 users. We crawled the README files using the official API of GitHub~\cite{githubapi}. Due to the rate limits for API requests, we choose only README files in projects with no less than 10 stars.\footnote{https://help.github.com/articles/about-stars/, retrieved in August 2018} Table~\ref{tab:dataset} summaries the data set.

Note that the data of emoji reaction,\footnote{\label{foot:reaction}https://github.com/blog/2119-add-reactions-to-pull-requests-issues-and-comments, retrieved in August 2018.} which is provided by GitHub since March 2016 to respond to conversational posts, are not used in this work. Instead of the six emojis (i.e., \emoji{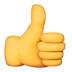}, \emoji{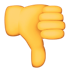}, \emoji{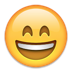}, \emoji{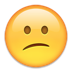}, \emoji{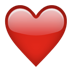}, and \emoji{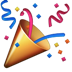}) provided by the reaction function, we collect the emojis that the users spontaneously typed into the free text for they have a greater variety in both the adopted type and usage pattern. %and are more representative of the communications. %One reason is that we are not able to obtain the timestamps of the reactions and can not attribute them to particular users. 

%Note that although after March 2016 GitHub users are able to respond to conversational posts using system provided emoji reaction,\footnote{\label{foot:reaction}https://github.com/blog/2119-add-reactions-to-pull-requests-issues-and-comments, retrieved in August 2018.} we are not able to obtain the timestamps, and they can not be attributed to particular users. Instead, emojis that the users spontaneously typed into the free text have a greater variety and are more representative of their personalities. Therefore, in this study, we only look at the emojis in the free text of the conversations. This also makes our findings more comparable to the results on other online platforms.

\begin{table*}[htbp]
\centering
\caption{Summary of Data Set.}
\begin{tabular}{c|rrrrr}
\hline
& \#issue & \#issue comment & \#pull request & \#pull request comment & \#README \\\hline
Non-emoji 	&	30,805,343 	&	49,173,170 	&	12,696,474 	&	23,950,062 	&	744,660 \\
Emoji 	&	60,742 	&	307,298 	&	34,576 	&	407,302 	&	21,444 \\\hline
Total 	&	30,866,085 	&	49,480,468 	&	12,731,050 	&	24,357,364 	&	766,104 \\\hline
\end{tabular}
\label{tab:dataset}
\end{table*}

\subsection{Emoji Popularity}
Before looking at the emojis, we measure their popularity by demonstrating the proportions of posts and users that adopted emojis in the studied period. We include only the conversational posts here because the README posts are not with timestamps.

%Statistics show that the fraction of emoji posts keeps increasing from nearly zero in 2012, while 0.58\% of issues, 0.75\% of pull requests, 1.32\% of comments for issues, and 3.23\% of comments for pull requests contain emojis in June, 2017. Although the proportion is much lower than the coverage of emojis on nontechnical communities such as Twitter (13.69\% as reported in \cite{pavalanathan2015emoticons})

The fraction of emoji posts remains at nearly zero in the first few years in our data set, increasing slowly, and shows a sharp increase since March, 2016. Statistics show that in June, 2017, 0.58\% of issues, 0.75\% of pull requests, 1.32\% of comments for issues, and 3.23\% of comments for pull requests contain emojis. Although the proportion is much lower than the coverage of emojis on nontechnical communities such as Twitter (13.69\% as reported in \cite{pavalanathan2015emoticons}) possibly due to the nature of technical discussions, the proportion keeps increasing over time in the studied period. 

\begin{figure}[hbt]
\center
  \includegraphics[width=0.95\columnwidth]{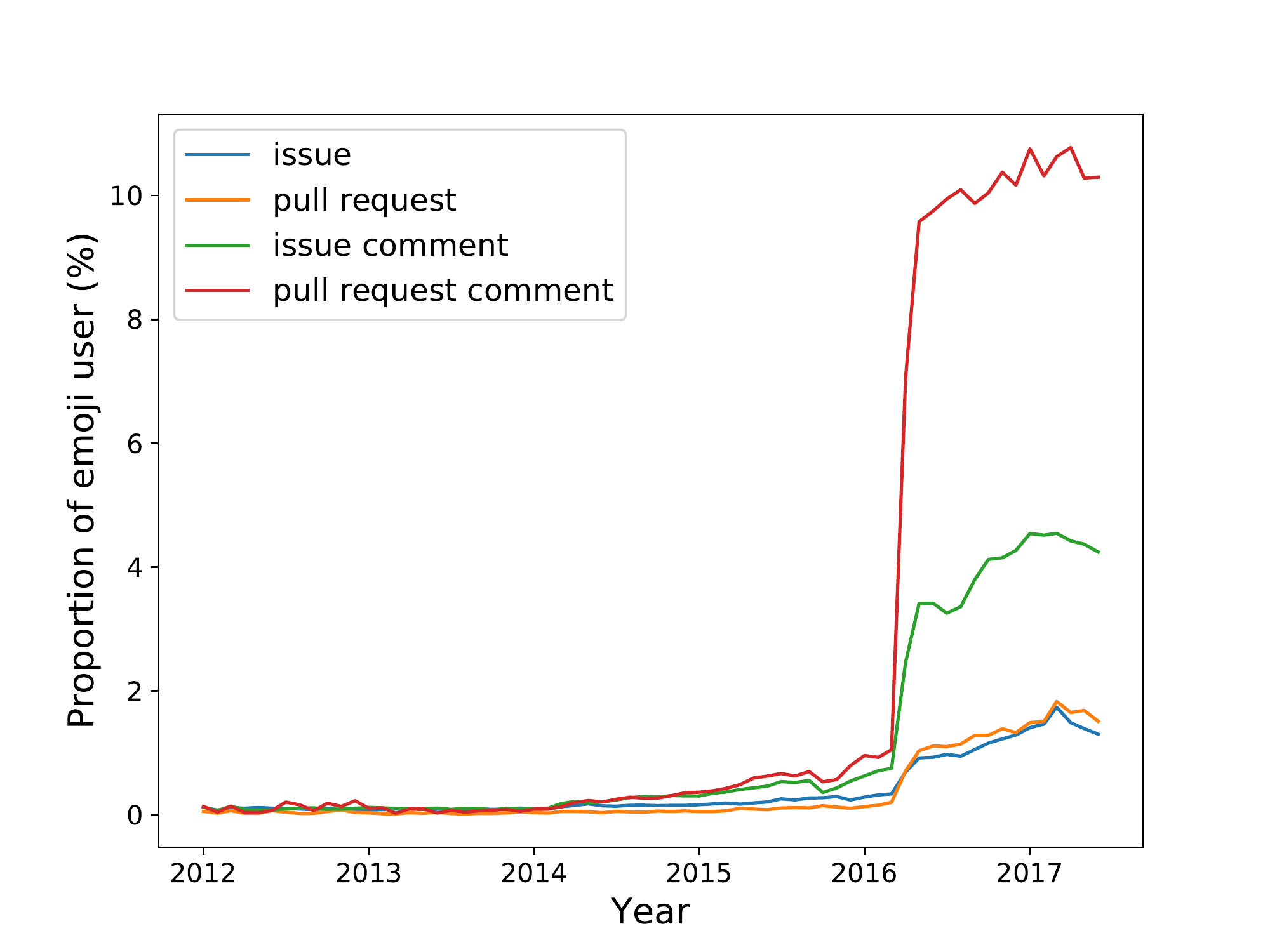}%{pics/emoji_usernum_prop}user-proportion
  \caption{Proportion of emoji users in GitHub conversations grows over the year 2016, with a sharply increase after the release of the emoji reaction feature.}
  \label{fig:proportion}
\end{figure}

We can find a quite high proportion of users who actively use emojis in their conversations. Figure~\ref{fig:proportion} illustrates the proportion of users who have used at least one emoji in each month, over all users involved in different types of posts in the same month. The proportion of emoji users started from below 1\% in the beginning of the year 2016 and increased over time. Users are more likely to use emojis in comments than in the original posts of issues and pull requests. In all four categories of the conversational posts, there was a significant increase of emoji users from March to April of 2016, which coincides with the release of the emoji reaction feature. Among users who commented on pull request, this proportion increased sharply for nearly seven times in April. 10.29\% of users who commented on pull requests in June of 2017 used emojis in their comments, a surprisingly larger proportion compared to the proportion of emoji posts. As the emoji reactions are not included, the proportion of emojis users may be underestimated. %As we are not able to attribute emoji reactions to users and to the months, the proportion of emoji users may be underestimated. 

\subsection{Emoji Distribution}\label{sec:dis}
%Before looking into the emoji posts, we explore a bit further to \textit{where are emojis more likely to be used} and \textit{who are more likely to use emojis}, with the support of our data set. 

Emojis are not evenly distributed in posts. For example, 0.20\% of the issues in our data set contain at least one emoji. After grouping the issues by the number of comments, we find the proportion of emoji issues is 0.40\% for the top 10\% issues, significantly higher than 0.20\% ($p<0.001$). Interestingly, such a doubled proportion can be observed in most months in our data set. This finding implies a positive relation between emoji usage and responses to issues.
%Recalling that only 0.58\% of issues in June of 2017 contain emoji, we find 1.20\% of the top 10\% issues with the most comments contain emojis. 

Similar findings can be made for the users. In general, 3.66\% of users have used at least one emoji in conversational posts. When restricting the scope of users to those who have posted issues, this proportion is significantly higher to be 4.71\%  ($p<0.001$). Further, we rank the users in descending order of number of issues they posted, and find that this proportion is 23.39\% for the top 10\% of the users, significantly higher than 4.71\% ($p<0.001$). Such findings indicate the potential correlation between emoji usage and user activeness.

Considering that the use of emojis can be influenced by multiple factors, we leave further research of the possible effect of using emojis in the participation and activeness of users for future work. In this paper we focus only on the posts with emojis.
%The use of emojis can be influenced by multiple factors, and can possibly influence the participation and activeness of users. We leave further research of this direction for future work and focus on understanding emojis in this paper.
 
%During the studied period, 3.66\% of users have used at least one emoji in their conversational posts. 

%\noindent $\bullet$ \textbf{Data cleaning}.

%\textbf{Privacy and Ethics.} We use public data collected from GitHub. Only discussions among developers are included in the study rather than codes or documents of GitHub projects. Personal identifiers are removed in the entire analysis. This work has been approved by the appropriate Institutional Review Board (IRB).

%% file: sections/characteristics.tex
\section{Usage Characteristics}\label{sec:characteristics}

%Although we have obtained the general interpretations of emojis on GitHub, the exact meaning and effect of emojis in a post can be influenced by how it is used in its context. This section addresses \textbf{\textit{RQ2: How are emojis used by developers on GitHub?}} by investigating the density, position, and selection of emojis on this platform. 

In this section we address \textbf{\textit{RQ1: How are emojis used by developers on GitHub?}} by looking at the top emojis and typical usage patterns on GitHub. 

\subsection{Top Emojis}\label{sec:pre}
In total, there are 1,271 emojis used on GitHub.
%The popular emojis can provide clues of the topics, sentiments, and atmosphere of a platform. 
Ranked by occurrence, the 10 most used emojis are \emoji{519}, \emoji{834}, \emoji{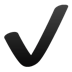}, \emoji{331}, \emoji{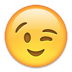}, \emoji{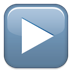}, \emoji{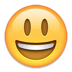}, \emoji{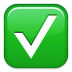}, \emoji{144}, and \emoji{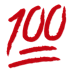}. To compare with, the 10 most used emojis on Twitter tracked by EmojiTracker\footnote{\url{http://emojitracker.com}, retrieved on September 3, 2017.} are \emoji{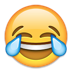} (21), \emoji{144} (9), \emoji{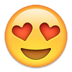} (57), \emoji{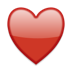} (41), \emoji{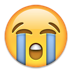} (47), \emoji{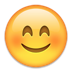} (11),  \emoji{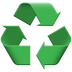} (222), \emoji{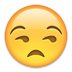} (123), \emoji{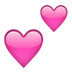} (375), and \emoji{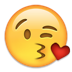} (152). The number in parentheses shows the ranking of the corresponding emoji on GitHub.  %Most of the 10 emojis are faces with emotions or hearts, and the left \emoji{79} means recycling. %As reported in literature~\cite{Lu:UbiComp16}, the 10 most used emojis in a popular input app called the Kika Keyboard were \emoji{832}, \emoji{144}, \emoji{843}, \emoji{840}, \emoji{61}, \emoji{518}, \emoji{874}, \emoji{875}, and \emoji{591}, sufficiently consistent to Twitter. 
Interestingly, only one emoji (\emoji{144}) is overlapped by the two sets of top 10 emojis. Most of the left 9 popular emojis on Twitter rank pretty low on GitHub. The most popular emoji \emoji{832} (\textit{face with tears of joy}) on Twitter, which was even elected as ``\textit{Oxford Dictionaries word of 2015}'',\footnote{\url{http://blog.oxforddictionaries.com/2015/11/ word-of-the-year-2015-emoji}} is the 21\textit{st} most used emoji on GitHub.

We conduct a \textit{Wilcoxon signed-rank test}~\cite{wilcoxon1945individual} on rankings of emojis on the two platforms and find significant difference ($p=0.002$ for top 10 emojis, $p=0.000$ for top 50 emojis), indicating domain specific preferences for emojis on this technical platform.

\begin{table}[htbp]
\caption{The Most Used Emojis.}
\begin{center}
\begin{tabular}{|c|c|}
\hline
\textbf{Post Type}& \textbf{Top 10 emojis} \\
\hline
All & \emoji{519}\emoji{834}\emoji{128}\emoji{331}\emoji{839}\emoji{35}\emoji{833}\emoji{119}\emoji{144}\emoji{617}  \\
\hline
issue & \emoji{128}\emoji{35}\emoji{519}\emoji{119}\emoji{834}\emoji{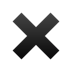}\emoji{144}\emoji{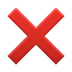}\emoji{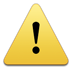}\emoji{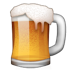} \\\hline
issue comment & \emoji{519}\emoji{834}\emoji{839}\emoji{128}\emoji{331}\emoji{833}\emoji{840}\emoji{35}\emoji{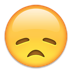}\emoji{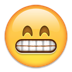} \\\hline
pull request & \emoji{519}\emoji{834}\emoji{128}\emoji{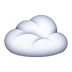}\emoji{331}\emoji{119}\emoji{89}\emoji{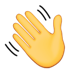}\emoji{839}\emoji{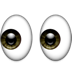} \\\hline
pull request comment & \emoji{519}\emoji{834}\emoji{331}\emoji{839}\emoji{128}\emoji{617}\emoji{35}\emoji{144}\emoji{944}\emoji{833} \\\hline
README & \emoji{128}\emoji{89}\emoji{119}\emoji{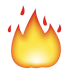}\emoji{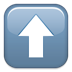}\emoji{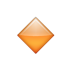}\emoji{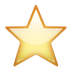}\emoji{519}\emoji{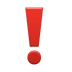}\emoji{144} \\\hline
%project description & \emoji{132}\emoji{121}\emoji{1005}\emoji{144}\emoji{76}\emoji{158}\emoji{328}\emoji{90}\emoji{507}\emoji{910} \\\hline
\end{tabular}
\label{tab:top}
\end{center}
\end{table}

In the top 10 emojis on GitHub, most emojis can be used to show positivity such as happiness, congratulations, praises, and appreciation, implying a positive atmosphere which can be of significance to communication and project collaboration. Another interesting observation is the popularity of \emoji{128}, \emoji{35}, and \emoji{119}, which can be used as ``bullets'' for a list of items.

When decomposed to different types of post (see Table~\ref{tab:top}), greater variety and more interesting usage can be found in the top emojis. For example, the ship (\emoji{944}), whose popularity ranks \textit{no}.609 on Twitter, is the ninth popular emoji in pull request comments. To understand the popularity of such emojis, we will interpret the meanings and intentions in the following sections.%We next introduce an embedding-based approach to understand the domain specific semantics of emojis on GitHub. % the ``bullets'' emojis (i.e., \emoji{128}, \emoji{35}, \emoji{119}, \emoji{129}, and \emoji{137}) holds half of the top 10 emojis in issues, with \emoji{128} as the most used emojis. Additionally, the popularity of some emoji such as the 

%Note that in most cases the check mark is not used to mean \textit{right} but often means that a task is \textit{finished} or the code has \textit{passed} some test, and the cross is not used to mean \textit{wrong} but \textit{failed}. The warning sign (\emoji{89}) is often used in issues, pull requests, and readme files to indicate that something needs attention. The ship (\emoji{944}), which is widely used in pull request comments, is a typically domain-specific emoji as it indicates the ``launch'' of something, such as a new project or a new feature. 
%\emoji{519}\emoji{834}\emoji{144}\emoji{89}\emoji{316}

%The top emojis on GitHub have already revealed domain-specific usage of emojis, which can be illustrated from two folds. On the one hand, a different set of emojis are preferred by users of GitHub in comparison with other platforms to help express ``customized'' topics and sentiments. On the other hand, some emojis have specific meanings that are different with their original definition or common use. We have manually checked some of the emojis with domain-specific meanings, yet it is quite time-consuming for all the 1,271 emojis used on GitHub. To this end, we next leverage an embedding-based technique to interpret the meanings of emojis on GitHub.

\subsection{Emojis in Text}\label{sec:intext}
Used as complements or surrogates of plain text, emojis can generally make texts more vivid, expressive, and easy to read. How frequently are emojis used and where are they put in a post? We investigate the \textit{density} and \textit{position} of emojis in English texts. %How are emojis used on GitHub to help content expression? To answer this question, we first look at the density (i.e., the number of emojis normalized to the length of post) of emojis in each post. Considering the different orientations, we calculate the density in each post type respectively.

\begin{figure}[hbt]
\center
  \includegraphics[width=0.9\columnwidth]{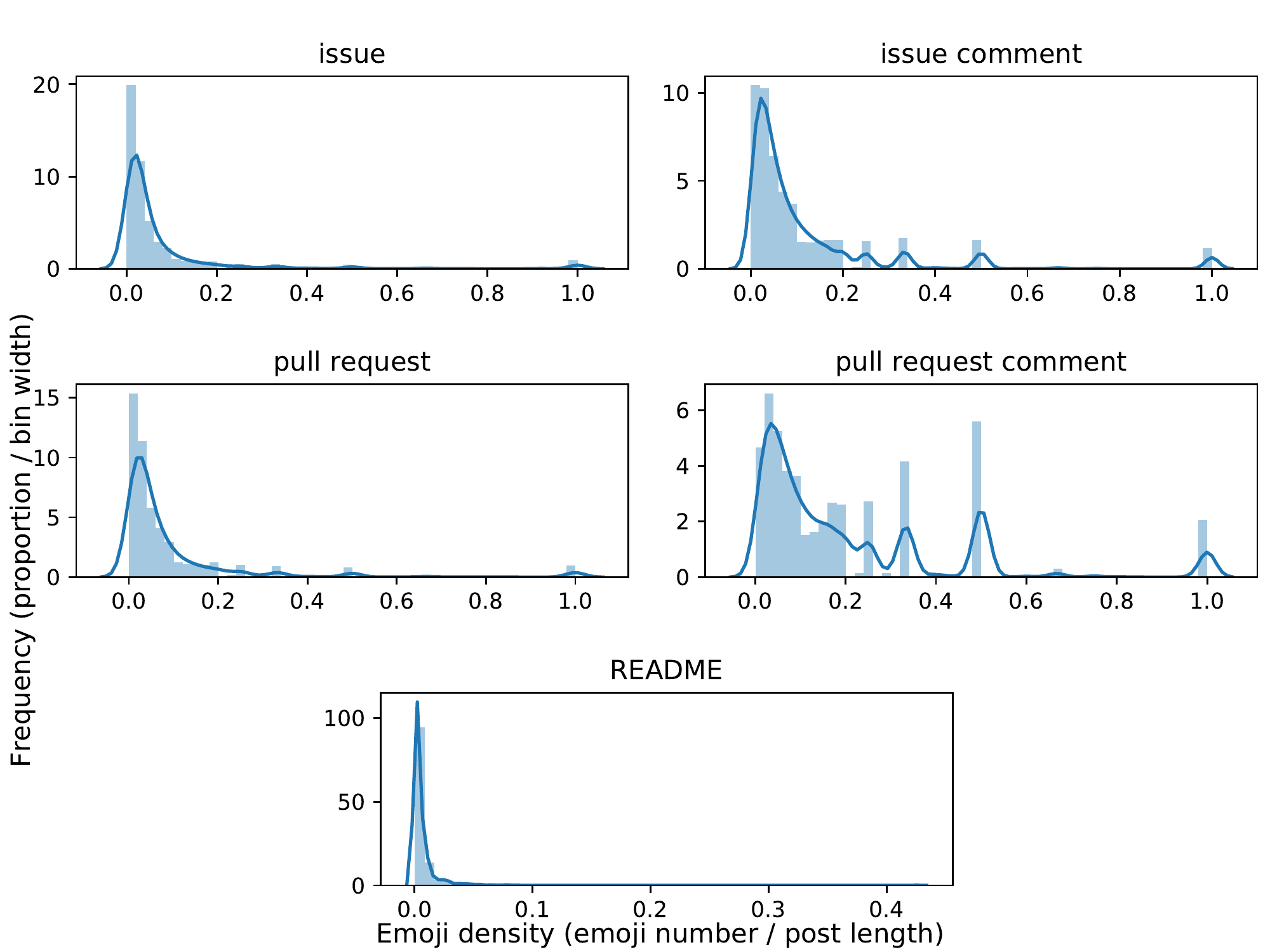}
  \caption{Emoji density in different posts.}
  \label{fig:density}
\end{figure}

\subsubsection{Density}
We define the density of emojis as the number of emojis normalized to the length of post, and present the density distribution with 
%The distribution of the emoji density is presented through 
frequency histograms in Fig.~\ref{fig:density}. Comparing the five distributions, we can find that in README the emoji density is particularly close to zero, possibly due to the relatively long texts. 
For the four conversational posts, the densities of emojis are mainly distributed between 0.0 and 0.2 with the largest peak near 0.0, indicating that emojis are mostly used with textual words, accounting for a small fraction of length in the post. Interestingly, there are discrete peaks with density of 0.25, 0.33, 0.5, and 1, especially in comments for issue and comments for pull request. We trace back and find massive emoji posts with 3 words (e.g., ``\textit{hooray for tests \emoji{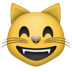}}''%\footnote{https://github.com/dimagi/commcare-hq/pull/12431\#issuecomment-232479922}
), 2 words (e.g., ``\textit{Whoops \emoji{831}. Merging!}''%\footnote{https://github.com/timehop/apns/pull/8\#issuecomment-65463719}
), 1 word (e.g., ``\textit{Thanks! \emoji{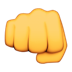}}''%\footnote{https://github.com/spite/rstats/pull/23\#issuecomment-232513405}
), and even no word (e.g., ``\textit{\emoji{519}}''%\footnote{https://github.com/ankurp/Dollar/pull/34\#issuecomment-46296465}
). Such use of emojis help explain the discrete peaks.

\subsubsection{Position}
Emojis tend to come at the end of messages, providing cues about how to understand the words that came before them~\cite{time2014emoji}. To verify this rule for emojis on GitHub, we extract emoji sentences from conversational posts and find most of them end with emojis. In particular, emojis are at the end of 66.77\% emoji sentences in issue comments.
%he density analysis indicates that emojis are mostly used with textual words. On Twitter, emojis tend to come at the end of messages~\cite{time2014emoji}, then what about GitHub? Considering the variety of the length of GitHub posts, we study the position of emoji from two levels of granularities, i.e., the sentence level and the post level.

%We extract the emoji sentences from the conversational posts, exclude non-textual ones, and map the position of emoji in a sentence to ten slots following Equation~\ref{equ:slot}, where $n$ represents the number of words in the sentence, $i$ represents the index of the emoji (counting from 1), and $\sigma$ is the smoothing factor. The heat map in Fig.~\ref{fig:sentence_pos} illustrates the distribution of emoji appearance probability. In the posts, emojis are mostly used in the end of sentences especially in issue comments (66.77\%). In addition, emojis also tend to occur in the beginning of sentences, indicating a pattern of showing stance or emotion with emojis before statements with texts.
%
%\begin{equation}
%	pos=\lfloor10\times\frac{i-1}{n-1+\sigma}\rfloor 
%	\label{equ:slot}
%\end{equation}
%
%\begin{figure}[hbt]
%\center
%  \includegraphics[width=1\columnwidth]{pics/sentence_pos_heatmap_4}
%  \caption{Emoji position in a sentence.}
%  \label{fig:sentence_pos}
%\end{figure}

\begin{figure}[hbt]
\center
  \includegraphics[width=1\columnwidth]{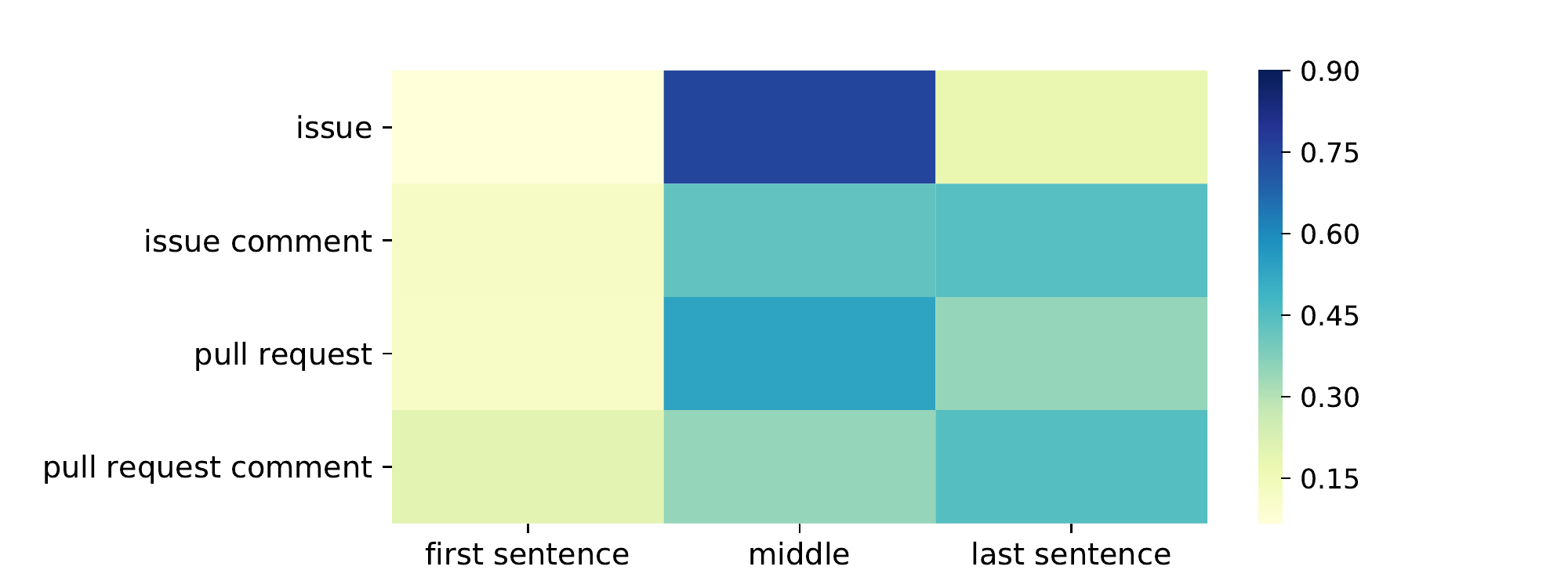}
  \caption{Emoji position in a post.}
  \label{fig:post_pos}
\end{figure}

At the post level, we segment sentences in a post to three position classes~\cite{Naaman:2018ce}, i.e., the first sentence, the last sentence, and the left in middle. The distribution of emojis in posts containing no less than three sentences is shown in Fig.~\ref{fig:post_pos}. It can be observed that emojis are mostly used in the middle of issues (75.34\%) and pull requests (53.83\%). However, in the two types of comments, emojis are more likely to be used at the end. Considering that users often express attitudes or emotions in comments, they can use emojis at the end of comments to enrich such expression.

%At the post level, we refer to~\cite{Naaman:2018ce} to map the sentences in a post to a three-class BEGIN/MID/END. Emojis in the first sentence will be grouped to the BEGIN class,  in the last sentence will be grouped to the END class, and the left to the MID class. As shown in Fig.~\ref{fig:post_pos}, for issues containing no less than three sentences, 75.34\% of emojis are used in the MID. In pull requests, the probability of containing emojis in MID position is also the highest, i.e., 53.83\%. However, in the two types of comments, emojis are more likely to be used in the END than the MID of the posts.

\subsection{Appearance Patterns}
When emojis appear in posts on GitHub, three typical patterns including using an emoji as a post, repeating the same emoji, and assembling different emojis, can be observed. We study such patterns and illustrate them with examples.%We study such scenarios and try to tease out the emoji selection patterns.

\noindent $\bullet$ \textbf{A single emoji as a post}. We are interested in the posts with only one emoji but no plain texts, because the understanding of such a post relies on the emoji. We suppose that emojis in such cases can convey relatively clear meanings, otherwise they will be resistance to the communication. Ranking emojis by their possibilities to independently constitute a post, we have \emoji{519} (26.33\%), \emoji{944} (23.76\%), \emoji{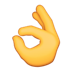} (16.67\%), \emoji{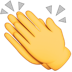} (16.13\%), \emoji{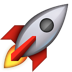} (15.23\%), \emoji{617} (13.24\%), \emoji{119} (12.50\%), and \emoji{331} (10.82\%) as the top ones. With our interpretation approach in Section~\ref{sec:interpretation}, the \emoji{944} and \emoji{910} can indicate the ``launch'' of something such as a new project or a new feature, and the rest ones can show attitudes or emotions. In contrast, emojis such as \emoji{89} and \emoji{35} are seldom used independently for they need supplementary words to express an complete idea.

%\noindent $\bullet$ \textbf{Single emoji as a post}. Some emojis are observed to be frequently used as a complete post without any other words. The most popular emoji \emoji{519} has the highest possibility (i.e., 26.33\%) to be a complete post, followed by \emoji{944} (23.76\%), \emoji{518} (16.67\%), \emoji{521} (16.13\%), \emoji{910} (15.23\%), \emoji{617} (13.24\%), \emoji{119} (12.50\%), and \emoji{331} (10.82\%).  Compared with the emojis least likely to be a complete post (such as \emoji{89} and \emoji{35} %, \emoji{129}, \emoji{128}), these emojis can have independent and clear meanings even without supplementary words.

%When multiple emojis are used in a post, there are two basic situations. One is that an emoji is repetitively used, and the other is that different emojis are co-used in one post.

\noindent $\bullet$ \textbf{Emoji repetition}. Emojis are often repetitively used to communicate a particular type of effect such as emphasis~\cite{jackson2016pragmatics}. %Similar to other platforms\reminder{cite}, emojis are often repetitively used on GitHub. Repetition can be regarded as a emphasis\reminder{cite}. 
On GitHub, \emoji{832}, \emoji{518}, \emoji{521}, \emoji{331}, and \emoji{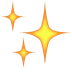} are with the highest possibilities to be used repetitively. %, yet not all emojis are used repeatedly. The emoji \emoji{832} has the highest possibility (i.e., 7.91\%) to be repeatedly used as \emoji{832}\emoji{832} or \emoji{832}\emoji{832}\emoji{832}, followed by \emoji{518} (5.20\%), \emoji{521} (4.37\%), \emoji{331} (2.26\%), and \emoji{132} (1.35\%). 
Interestingly, these emojis tend to convey positive attitudes or sentiments in most cases. For example, the comment ``\emoji{521}\emoji{521}\emoji{521}'' in an issue comment%\footnote{\url{https://github.com/cozy/cozy-drive/pull/48#issuecomment-276309446}}
 is to express confirmation and praise to the issue assignee, after which the corresponding commit is merged to master. In contrast, emojis that are the least likely to used in repetition include \emoji{89}, \emoji{129}, \emoji{851}, \emoji{520}, and \emoji{860}, which often mean warning, failure, disappointment, disagreement, and sadness. Such finding implies a friendly atmosphere on GitHub, which confirms the observation in Section~\ref{sec:pre}.
%Such finding implies the tendency of GitHub users to emphasize positive attitudes and sentiments instead of negative ones, which confirms the positive atmosphere discussed in Section~\ref{sec:pre}. 
%Some emojis tend to appear repetitively. For example, when occur in a post, the emoji \emoji{832} has a 7.91\% possibility to have the appearance of \emoji{832}\emoji{832} or \emoji{832}\emoji{832}\emoji{832}. Following are \emoji{518} (5.20\%), \emoji{521} (4.37\%), \emoji{331} (2.26\%), and \emoji{132} (1.35\%). On the other side, the emojis least likely to repeat are \emoji{89}, \emoji{129}, \emoji{851}, \emoji{520}, and \emoji{860}. Such observation implies a tendency to express positive attitudes and sentiments with emphasis but not to emphasize negative attitudes or sentiments.

%As an analogy to textual words~\cite{Islam:2018hz}, repetitive emojis can be considered sentimental in some cases. 

\noindent $\bullet$ \textbf{Emoji assembling}. In some cases emojis are assembled to express a complete and even complicated meaning in a post. %Emojis can occur with other emojis in one post. 
For example, \emoji{617} often occurs together with \emoji{519}, \emoji{518}, \emoji{506}, \emoji{331}, or \emoji{132}. %Emojis that most likely to occur together with other emojis are \emoji{617} (often co-occur with \emoji{519}, \emoji{518}, \emoji{506}, \emoji{331}, or \emoji{132}), \emoji{518} (often co-occur with \emoji{506}, \emoji{617}, \emoji{519}, \emoji{832}, or \emoji{144}), 
One typical example of such co-occurrences comes from an issue comment saying ``\textit{\emoji{506}\emoji{617}}''.%\footnote{https://github.com/banjoman/brilliant-bash/pull/15\#issuecomment-276224887}
~The writer of this comment is the owner of the project and he merged the corresponding commit after the comment. The two emojis are combined to make up a sentence, which can be inferred to mean ``\textit{I have reviewed this commit, it's perfect!}'' 
Another example is ``\textit{Thanks for this! We're checking it out. \emoji{518}\emoji{506}\emoji{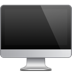}}''%\footnote{https://github.com/duckduckgo/zeroclickinfo-spice/pull/1983\#issuecomment-117684230}
~in an issue comment. The comment writer, a member of the project, added a label to the commented issue and reviewed the committed code after the comment. In this example, the three emojis are combined to retell the meaning of the plain texts, making the expression more vivid. %express a meaning, which has been expressed by the plain texts. 

Understanding assembled emojis may be not as easy to understand single or repeated ones, yet such usage can spice up the communication as the assembling can be quite creative especially when combined with domain knowledge or slangs. For example, in the README of the project Releasor,%\footnote{https://github.com/kimmobrunfeldt/releasor}
~i.e., ``\textit{\emoji{308} \emoji{463} Releasor is used in ProgressBar.js, git-hours, arr-mutations, and many others}'', the \textit{dog food} indicates that the Releasor tool has been used in the developers' own projects and demonstrates confidence.\footnote{\textit{Eating your own dog food}, or \textit{dogfooding}, is a slang term used to refer to a situation in which an organization uses its own product. This can be a way for an organization to test its products in real-world usage. Hence dogfooding can act as quality control, and eventually a kind of testimonial advertising. See https://en.wikipedia.org/wiki/Eating\_your\_own\_dog\_food, retrieved in August, 2018.}

\noindent \textbf{Summary}. 
This section characterizes emoji usage on GitHub from the aspects of the top emojis, the density and position of emojis in posts, and typical appearance patterns. We find that the favored emojis on GitHub are quite different from Twitter, indicating domain-specific usage such as technical discussions. In comments for issues and pull requests, emojis are often co-used with few words, and often come at the end of long comments. Additionally, emojis show different potentials to independently constitute a post, to emphasize sentiments by repetition, and to illustrate complicated meanings by assembling. It should be noted that these findings are not rules but usage patterns learned from data.

%answers RQ1 by investigating \textit{how often to use emoji}, \textit{where to put emoji}, and \textit{which emojis to choose} in GitHub communications. We find that in general the amount of emojis in a post is rather small in comparison with the length of the post, meanwhile emojis are often used in short texts with several or even no words in comments for issues and pull requests. We also find that  similar to Twitter, emojis often come at the end of a sentence. They are frequently used in the middle of an issue or a pull request, while more likely to be in the ending sentence of a comment. Additionally, emojis show different potentials to independently constitute a post, to emphasize sentiments by repetition, and to illustrate complicated meanings by assembling. It should be noted that these findings are not rules but usage patterns learned from data. To further understand the patterns and the developers, we next study the intentions of using emojis through a manual annotation task.

%% file: sections/interpretation.tex
\section{Emoji Interpretation}\label{sec:interpretation}
Although emojis are regarded as a ubiquitous language across different countries and user groups~\cite{Lu:UbiComp16}, we propose that the interpretation of emojis, similar to textual words~\cite{Islam:2018hz}, would have specific meaning in a technical field. In fact, we have obtained some clues for domain-specific usage of emojis. To address this research question (i.e., \textbf{\textit{RQ2: Do emojis have domain-specific meanings by developers on GitHub?}}), we develop an embedding-based approach to interpret the meanings of emojis and study the sentiment distribution of emojis on GitHub. %we first compare the most popular emojis on GitHub with a typical nontechnical platform (i.e., Twitter) to find domain-specific clues, then develop an embedding-based approach to interpret the meanings of emojis on GitHub.

\subsection{Semantic Understanding}\label{sec:semantics}

%RQ1.2: \textbf{\textit{Can the semantics of emojis interpreted automatically?}}

Developers on GitHub have been spontaneously proposing standards for using emojis to fit the functions and scenarios in this community. For example, \textit{gitemoji}\footnote{https://github.com/carloscuesta/gitmoji} provides an initiative to standardize and explain the use of emojis on GitHub commit messages. However, \textit{how it is defined} does not necessarily determine \textit{how it is used}. An example is that the \emoji{331} defined as \textit{party popper} in Unicode\footnote{http://unicode.org/emoji/charts/full-emoji-list.html} represents \textit{initial commit} in \textit{gitemoji}, yet this emoji is often used to express congratulations. 

%Such definition efforts suggest that emojis have localized meanings to fit the functions and scenarios in this community. 

We decide to study the interpretations of the emojis by developers on GitHub with the state-of-the-art text representation learning methods. By projecting language tokens into a semantic space, we are able to directly assess the meanings of emojis.
%The free texts in \textit{issues}, \textit{issue comments}, \textit{pull requests}, and \textit{pull request comments} are in format of markdowns. 
We extract all emojis and English words from the markdown\footnote{https://guides.github.com/features/mastering-markdown/} texts, replace each code block with ``[code]'' and each URL with ``[url]'', and filter out punctuation and special characters. We use the NLTK package\footnote{\url{http://www.nltk.org}} to tokenize and stem the words.

With the processed texts, we use word2vec~\cite{mikolov2013distributed} to train a 300-dimension embedding for each token, including words and emojis. Such a semantic space reflects the developers' interpretation of the meanings of both words and emojis. Based on cosine similarities between the embedding vectors, we are able to find the closest neighbors of any emoji in the semantic space. These neighbors, either words or emojis, can help us infer the meaning of the target emoji. %We give two examples of typical emojis, \emoji{469} and \emoji{944}, on GitHub.
We rank the closest word tokens to the given emoji in the semantic space. To compare, we also report the closest neighbors of the emoji in a different semantic space,\footnote{http://sempub.taln.upf.edu/tw/emojis/} in which the embeddings of emojis were trained with 10 millions Tweets posted by USA users~\cite{barbieri2016does}. This alternative semantic space represents the interpretations of emojis in common Internet communications.%by the common Internet users.

\newcommand{\tabincell}[2]{\begin{tabular}{@{}#1@{}}#2\end{tabular}}  
\begin{table}[htb]
\begin{center}
\caption{Similar Words of Emojis Based on word2vec Embeddings.}
\begin{tabular}{|c|c|c|} 
\hline
 & Similar words on GitHub & Similar words on Twitter %& Similar emojis on GitHub & Similar emojis on Twitter
 \\\hline
 \emoji{469} & \tabincell{c}{snuck, elus, insidi, uncov, \\nonbreak, \textbf{bug}, untitl, skater, \\crept, \textbf{smell}, \textbf{worrisom}, \textbf{fix}, \\\textbf{report}, gonzal, buggi, glare, \\floater, badminton, undetect, \\spectacular, miscellan} %full period
 %\emoji{469} & \tabincell{c}{\textbf{sorry}, snuck, elus, resurfac, \textbf{misalign}, dupe, \\ufa, \textbf{misbehavior}, \textbf{glitch}, spurious, ouch, \textbf{bug}, \\\textbf{odd}, peski, not, \textbf{weird}, \textbf{bizarr}, legit, \\askew, linger} %2016
 & \tabincell{c}{luvbug, Modjeska, zoomars, natie, \\getstheworm, buzzzzz, superfun, \\pangderbear, grindcore, kcm, \\localis, arcane, royalbluegroceryhp, \\coulby, meow, hunnies, gyp, \\kymimim, perdy, djcdannylewis} 
 %& \tabincell{c}{\emoji{472}, \emoji{470}, \emoji{737}, \emoji{736}, \emoji{570}, \\ \emoji{986}, \emoji{581}, \emoji{862}, \emoji{18}, \emoji{928}}
 %& \tabincell{c}{\emoji{472}, \emoji{454}, \emoji{470}, \emoji{797},  \emoji{1010}, \\ \emoji{254}, \emoji{476}, \emoji{576}, \emoji{502}, \emoji{482}} 
 \\\hline
 %\emoji{944} & \tabincell{c}{\textbf{merge}, \textbf{lgtm}, mayday, champagn, bruis, \\hurrah, \textbf{land}, leisure, woo, nit, hound, \\lox, moratorium, \textbf{approve}, \textbf{pass}, review, \\\textbf{resubmit}, squash, ahoy, mourner} %2016
 \emoji{944} & \tabincell{c}{\textbf{merg}, \textbf{lgtm}, \textbf{ship}, readi, \\nit, \textbf{approv}, \textbf{land}, leisur, \\ahead, squash, alright, ahoy, \\hooray, tum, thanksgiv, \\punt, immin, thatch, \\hurrah, appeas, sucker}
 & \tabincell{c}{cruise, carnival, indiaday, \\disneycruiseline, Bahamas, \\norweigian, ship, norweigan, \\jaymullins, Norwegian, \\jordanknight, sail, luby, maddecent, \\ncl, satsko, oob,  guana} %&\tabincell{c}{\emoji{612}, \emoji{331}, \emoji{320}, \emoji{467}, \emoji{1005}, \\ \emoji{328}, \emoji{672}, \emoji{216}, \emoji{323}, \emoji{906}}
% & \tabincell{c}{\emoji{120}, \emoji{634}, \emoji{373}, \emoji{227}, \emoji{934}, \\ \emoji{613}, \emoji{977}, \emoji{668}, \emoji{208}, \emoji{784}}
 \\\hline
\end{tabular}
\label{tab:similar}
\end{center}
\end{table}

Following to the word2vec results, in Table~\ref{tab:similar} we report similar words of two typical emojis, \emoji{469} and \emoji{944}, on GitHub.
For the emoji \emoji{469}, the most similar words on GitHub are obviously different with those on Twitter. For example, the similar words on GitHub includes \textit{bug}, \textit{smell}, \textit{worrisom}, \textit{fix}, and \textit{report}.
%\textit{sorry}, \textit{misalign}, \textit{misbehavior}, \textit{glitch}, \textit{bug}, \textit{odd}, \textit{weird}, and \textit{bizarr} (stem of bizarre). 
Such word neighbors reflect the domain specific meaning of \emoji{469} on GitHub. That is, \emoji{469} represents \textit{bugs} in code on GitHub, but not on platforms such as Twitter.

Similar conclusions can be derived for the emoji \emoji{944}. On Twitter, this emoji appears in similar contexts with words like \textit{cruise}, \textit{ship}, and \textit{sail}. However, on GitHub, it has similar meanings with words like \textit{merge}, \textit{lgtm} (look good to me), \textit{ship}, \textit{approv}, and \textit{land}. 
%\textit{merge}, \textit{lgtm} (look good to me), \textit{land}, \textit{approve}, \textit{pass}, and \textit{resubmit}. 
It indicates a status that the code is ready for use (to be ``shipped'').

In addition to the closest words to an emoji, we can also discover its closest emojis in the semantic space. For example, given that \emoji{469} is referred to as a code bug rather than an animal bug, one may expect that \emoji{469} has different emoji neighbors on GitHub than on Twitter. The 10 emojis that have the most similar embeddings with \emoji{469} on GitHub are \emoji{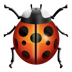}, \emoji{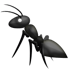}, \emoji{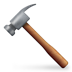}, \emoji{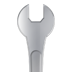}, \emoji{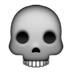}, \emoji{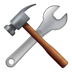}, \emoji{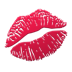}, \emoji{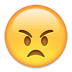}, \emoji{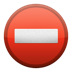}, and \emoji{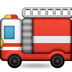}. In comparison, the 10 closest emojis on Twitter are \emoji{472}, \emoji{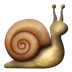}, \emoji{470}, \emoji{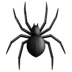}, \emoji{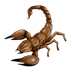}, \emoji{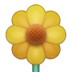}, \emoji{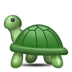}, \emoji{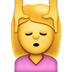}, \emoji{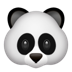}, and \emoji{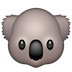}. The most interesting differences are the existence of \emoji{737}, \emoji{736}, and \emoji{986}, which point to the meanings of \textit{debugging} or \textit{fixing issues}.

%With the embeddings trained from the contexts of emojis, emojis on GitHub can be interpreted more accurately. To facilitate further research, we plan to release the embeddings and neighbors of emojis when this paper is published. 

\subsection{Sentiment Distribution}\label{sec:senti}
The observation of domain-specific preference for emojis and the implicated positive atmosphere motivate us to look at the sentiments expressed by emojis on this platform. By comparing the sentimental emojis on GitHub and Twitter may help us understand the culture of the developer community.

Given that emojis can have domain-specific meanings rather than their original definitions or interpretations by the public, their sentiments may also change. Instead of directly using the reported sentiments of certain emojis, we carefully extract the sentiment score of each emoji based on the sentiments of its neighboring words. 

Specifically, we calculate the sentiment score for each word with the SentiStrength-SE tool~\cite{Islam:2018hz}, which was designed for sentiment analysis in the software engineering domain. Each word has a score in \{-100, 0, 100\}, corresponding to \{\textit{negative}, \textit{neutral}, \textit{positive}\}. %with the LIWC package,\footnote{\url{http://liwc.wpengine.com}.} a value between -100 (the most negative) to 100 (the most positive). 
The sentiment score of a given emoji%\footnote{The embeddings, neighbors, and sentiment scores of emojis on GitHub will be released along with the paper's publication.}
~is the weighted average of the sentiment scores of its 100 nearest neighboring words, weighted by the cosine similarity between the embeddings of the emoji and the word. 

\begin{figure}[hbt]
\center
  \includegraphics[width=1\columnwidth]{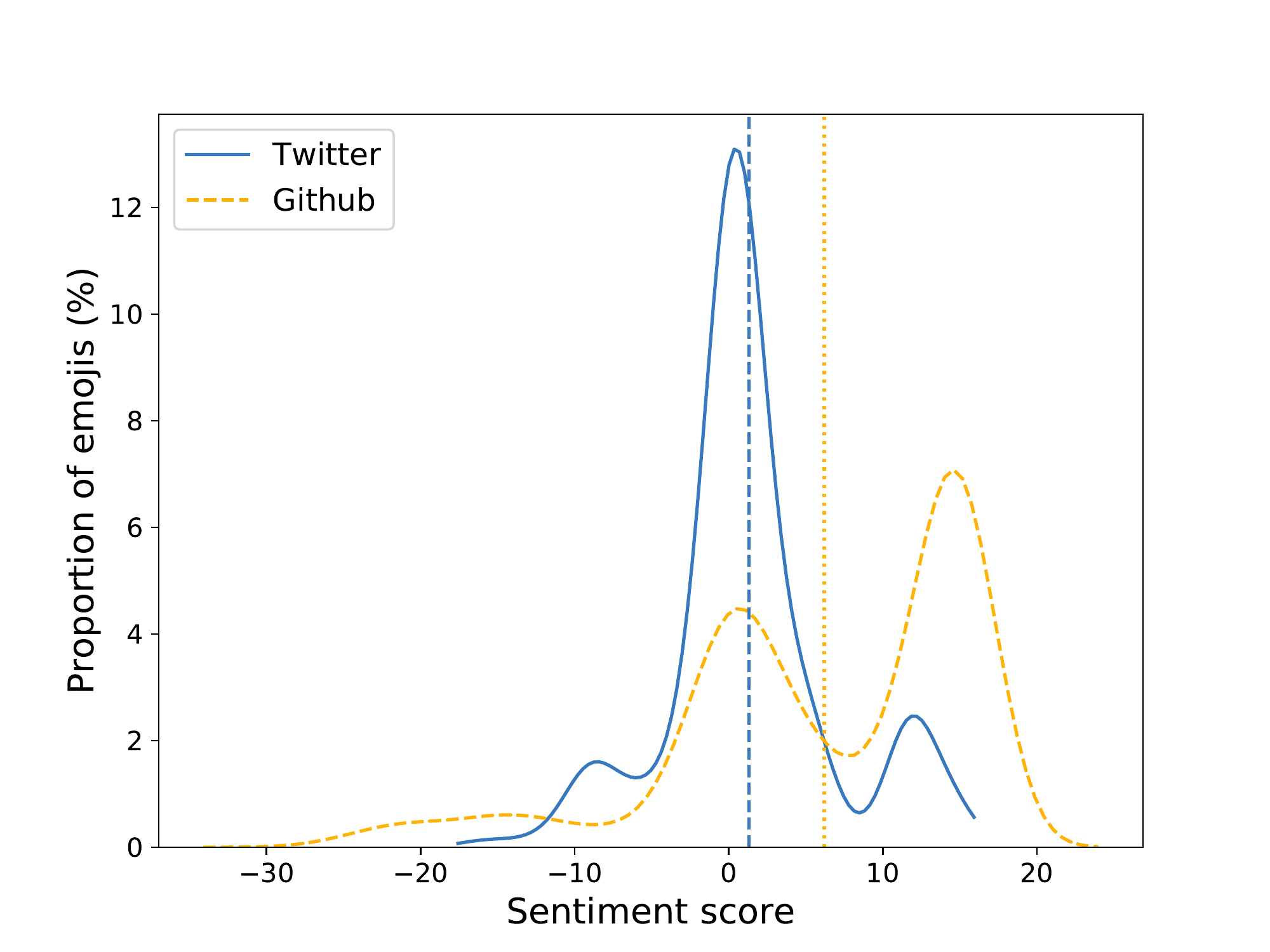}
  \caption{Sentiment distribution of emojis.}
  \label{fig:sentiment}
\end{figure}

After deriving the sentiment scores of emojis used on GitHub, we group the emojis into different score intervals and plot the total frequency of emojis used in GitHub posts against the sentiment scores in Fig.~\ref{fig:sentiment}. %The distribution demonstrates three interesting regions of sentiments, i.e., the positive, the neutral, and the negative sentiments. In specific, most emojis used in GitHub conversations are with positive emojis. Such observation suggests that programmers on GitHub tend to express positive sentiments through emojis to have better communications with others.
The distribution has a clear tendency towards positive sentiments, which suggests that programmers on GitHub tend to express positive sentiments through emojis, as support or appreciation for each other.

To understand the difference between developers and common Internet users, we also measure the sentiment scores of emojis used on Twitter and plot the same distribution (the frequency of emojis used on Twitter are obtained through Emojitracker). Sentiment scores of neighboring words of emojis on Twitter are calculated with the LIWC package.\footnote{\url{http://liwc.wpengine.com}} Clearly, the entire distribution of Twitter emoji sentiment shifts to the left ($p<0.001$ for the difference of mean), which suggests that emojis are more frequently used to express positive sentiments on GitHub compared to those on Twitter. 

%In summary, the descriptive analysis of emoji usage on GitHub reveals that contradicting to their common stereotypes, developers on GitHub use a large variety of emojis in their programming-related conversations. Programmers of different languages do have different tendency and preferences of using emojis, and the distribution of emojis they used is able to distinguish the different cultures of programming language communities. Comparing to the common Internet users, developers have their own interpretation of certain emojis, and they tend to use more positive emojis. Contradicting to the common stereotype of programming tasks as tedious and boring, the developer community presents a positive and supportive culture compared to the public. 

\noindent \textbf{Summary}. 
Analysis in this section evidences that emojis can have domain-specific interpretations by developers on GitHub in comparison with the common Internet users. Hence, the technical context should be considered for accurate understanding of emojis before leveraging emojis for further research. In addition, the developer community tend to use more positive emojis, which presents a positive and supportive culture compared to the public. %the domain-specific usage of emojis from two folds. On the one hand, the preference for emojis of GitHub users is differentiated from Twitter, possibly due to the nature of technical communications and the atmosphere of open source collaboration. On the other hand, emojis can show specific meanings, which should be interpreted in the technical context for accurate understanding. 
%The top emojis on GitHub have already revealed domain-specific usage of emojis, which can be illustrated from two folds. On the one hand, a different set of emojis are preferred by users of GitHub in comparison with other platforms to help express ``customized'' topics. On the other hand, some emojis have specific meanings that are different with their original definition or common use. We have manually checked some of the emojis with domain-specific meanings, yet it is quite time-consuming for all the 1,271 emojis used on GitHub. To this end, we next leverage an embedding-based technique to interpret the meanings of emojis on GitHub.

%% file: sections/purpose.tex
\section{Intention Understanding}\label{sec:intention}

%As nonverbal cues, emojis are universally used in online communications. Facial expressions, body gestures, animals, objects are used in conversations, making the communication more vivid. 
Intentions of emoji usage in daily communication have been studied~\cite{Hu2017Spice} while the understanding of intentions on a technical platform such as GitHub is lacked. In this section, we address the research question \textbf{\textit{RQ3: What are the intentions of using emojis by developers on GitHub?}} by proposing a taxonomy of intentions with a manual annotation task. %To give a 95\% confidence level and a 5\% confidence interval, we randomly select 400 posts in English for each of the five post types for annotation. %By analyzing the annotated intentions in different posts, \reminder{we find that}

%From the manual annotation process, we find that  understanding of emoji purpose can not be confined to the nearest sentence, instead, a wider range of contexts should be considered including the post and even the entire conversation.

\subsection{Taxonomy}
Based on existing studies about emoji usage intentions, we develop an initial set of intention categories and adapt them to emoji usage on GitHub with a subset of posts. The final taxonomy of intentions is as follows.
%Revised during the annotation process, the taxonomy of intentions is as follows. %Two of the authors manually annotated 500 randomly selected posts (100 posts for each kind), and discuss any discrepancies until they agree. The taxonomy is revised according to the annotation and discussion process, which is as follows.
%\begin{table*}[htbp]
%\caption{Taxonomy of Emoji Purposes}
%\begin{center}
%\begin{tabular}{|c|c|c|}
%\hline
%\textbf{Index} & \textbf{Category} & \textbf{Example} \\\hline
%1 & Sentiment & \\\hline
%1-1 & express sentiment & \textit{\emoji{519}Nice work!}\footnote{\url{https://github.com/sammanthp007/php-input-output-sanitization/issues/1}}\\\hline
%1-2 & weaken sentiment & \\\hline
%1-3 & strengthen sentiment & \textit{(that's right, it's blurring the actual content of the table cell! \emoji{871}}\footnote{\url{https://github.com/Khan/SwiftTweaks/issues/74}} \\\hline
%1-4 & reverse sentiment & \\\hline
%2 & Statement & \textit{This should not happen, I would even go as far as to call it a \emoji{469}.}\footnote{\url{https://github.com/hyperoslo/Spots/issues/407}}\\\hline
%3 & Content organization & \\\hline
%3-1 & Bullets, symbols & \\\hline
%3-2 & Eye-catching & \\\hline
%4 & Unintentional & \\\hline
%5 & Emoji & \\\hline
%6 & Conversational norm & \\\hline
%\end{tabular}
%\label{tab:taxonomy}
%\end{center}
%\end{table*}

%% a description, help understand the status

%% label, classification (propose a classifier?), can help to recognize posts with sentiments, facilitate sentiment analysis (how?)

%% help understand developers through their emoji usage

%% facial expression

\begin{enumerate}
	\item \textbf{Sentimental usage}. Using emojis can help express sentiments of users including their feelings, emotions, and attitudes. Considering the final effect of the expression (with the emoji) and the original sentiment of the plain text (without the emoji), the sentimental intention of using emojis can be de-composed to two sub-categories, i.e., sentiment expression and sentiment strengthening. Note that we also considered \textit{sentiment weakening} and \textit{sentiment reversing} at first, yet no such examples have been found. We claim this intention category can be extended if necessary.%Sentimental effects of emojis have been a typical reason for using them.  
	\begin{enumerate}
		\item Emojis conveying sentiment in a non-sentimental textual context have the intention of \textbf{sentiment expression}. For example, in the issue \textit{``Even the Guardian has TLS now.... \emoji{519}''}%\footnote{\url{https://github.com/novaramedia/novaramedia-com-v1/issues/3}}
		, the attitude of appreciation is expressed through the \emoji{519} emoji. %``\textit{\emoji{519}Nice work!}''\footnote{\url{https://github.com/sammanthp007/php-input-output-sanitization/issues/1}}%When there are no sentimental cues in the plain text while the emoji expresses some sentiment, the intention of using this emoji is \textbf{sentiment expression}. 
%		\item Sentiment weakening
		\item In the case that the plain text has already expressed some sentiment, the emoji often makes the post more expressive. We name this intention as \textbf{sentiment strengthening}. For example, in the issue ``\textit{(that's right, it's blurring the actual content of the table cell! \emoji{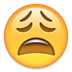}}'',%\footnote{\url{https://github.com/Khan/SwiftTweaks/issues/74}} 
		~the negative emotion is expressed by the text and strengthened by the \emoji{871} emoji.%comment ``\textit{Got it, thanks a lot for the help :) \emoji{617}}''\footnote{\url{https://github.com/Yalantis/FastEasyMapping/issues/79#issuecomment-282543620}}, 
%		\item Sentiment reversing 
	\end{enumerate}
	\item \textbf{Statement enriching}. Emojis can make expressions, which are not limited to sentimental expressions, more vivid. With this intention, emojis are often used to replace or illustrate contents such as concepts and objects in text. For example, in the issue  ``\textit{This should not happen, I would even go as far as to call it a \emoji{469}.}'',%\footnote{\url{https://github.com/hyperoslo/Spots/issues/407}}
	~the emoji \emoji{469} enriches this statement by representing the word \textit{bug}.
	\item \textbf{Content organization}. The pictographic nature of emojis makes them a good choice to assist the organization of contents in a post for readability improvement. When multiple items are listed, emojis such as \emoji{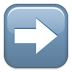} and \emoji{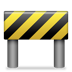} can be item bullets as alternatives of symbols like $\bullet$ and $\blacktriangleright$.  Additionally, some emojis such as \emoji{119} and \emoji{137} can be used in checklists with the semantics they conveyed.%\footnote{\url{https://github.com/keepassxreboot/keepassxc/pull/317}}  %A post organized with these emojis can be more methodical and readable.
	\item \textbf{Content emphasis}. To avoid being overwhelmed by massive content, the important points can be demonstrated with emojis to attract more eyeballs. Emojis with this intention are not necessarily semantically connected with the text but contribute with their pictographic characteristics. Typical emojis often used with such intention include \emoji{89}, \emoji{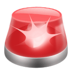}, \emoji{132}, and \emoji{158}.
	\item \textbf{Atmosphere adjustment}. Two main scenarios are included in this intention category. First, emojis can be used to adjust tone, making the messages less serious and more friendly. An example is ``\textit{Please do not be terrified \emoji{886}}'' in a pull request.%\footnote{https://github.com/chef/chef-server/pull/1241}
	%An example is the issue ``\textit{Not sure how this got into master... \emoji{836}}''.\footnote{\url{https://github.com/agenda-item/agenda-item/issues/116}} 
	As nonverbal cues, emojis especially facial expressions can be used with this intention. Although facial emojis are often combined with emotion, they do not show obvious emotion but often politeness and kindness in its context when categorized to this intention. Second, when one intends to say something but has no idea of a specific topic, they can use emojis, which can even be not semantically related to the context. %For example, the cherry emoji \emoji{276} is used as a comment\footnote{https://github.com/spatie/vue-save-state/pull/2\#issuecomment-270016279} to a pull request, possibly implying that the commit has been merged to the master and the work finished. 
	Emojis include hearts, gestures, animals, and objects can often be with this intention.%\footnote{\url{https://github.com/githubteacher/cautious-lamp/issues/1}}
	\item \textbf{Unintentional usage}. Emojis can be unintentionally used in posts when they are not input by the writer. Emojis in pasted contents including codes and logs are categorized to this intention.  
	\item \textbf{Emoji}. In some cases, emojis are just emojis, e.g., ``\textit{\emoji{154} is for upgrading dependencies, from other libraries, where  \emoji{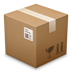} is for things like binaries, from the local project.}''%\footnote{https://github.com/carloscuesta/gitmoji/issues/61\#issuecomment-262525167}
\end{enumerate}

%In our taxonomy of intentions of using emojis, the 

\subsection{Annotation}
Based on the proposed taxonomy, we manually annotated the intention of using emojis in 2,000 emoji posts. To give a 95\% confidence level and a 5\% confidence interval, we randomly select 400 posts in English for each of the five post types for annotation. For multiple appearances of emojis in a post, only the first one is annotated except that it is used in combination with others (e.g., ``\textit{\emoji{949}~Work in Progress~\emoji{949}}''%\footnote{\url{https://github.com/TeamAlphaChess/Chess-app/pull/49}})
. We discuss the annotation process in detail before reporting the results.

\subsubsection{Discrepancies}
Understanding the meaning beyond language is often subjective and has always been challenging, where discrepancies can occur when different individuals interpret the same expression. The two authors who annotate the intentions of using emojis complete their task independently and discuss about the discrepancies until they agree. For example, in the statement ``\textit{I've tried to reinstall the game, but the error keeps happening \emoji{860}}'' in an issue,%\footnote{\url{https://github.com/ValveSoftware/csgo-osx-linux/issues/1345}} 
~one of the two authors classified the intention of \emoji{860} to \textit{sentiment strengthening} while the other believe it is to express an sentiment. After discussion, they agreed that the text ``\textit{I've tried to reinstall the game, but the error keeps happening}'' is stating a fact without obvious sentiments, but the emoji \emoji{860} brings emotion of sadness and disappointment to this expression; hence it is finally categorized to \textit{sentiment expression}. 

\subsubsection{Multiple intentions}
In some cases an emoji can not be categorized to only one intention. For emojis showing complicated intentions, we allocate multiple labels to them and select a primary intention for further study. For example, in the sentence ``\textit{\emoji{950}Please review the  guidelines for contributing to this repository.}'' of a pull request,%\footnote{\url{https://github.com/angular-translate/angular-translate/pull/1558}} 
~the \emoji{950} is effective in drawing attention (i.e., the \textit{content emphasis} intention) to this warning. Meanwhile, this \textit{police car light} emoji contains the semantic of emergency that enriches the context of warning, indicating an effect in making the statement more expressive. We label this emoji in this post with a \textit{content emphasis} intention for its main effect. Another example is ``\textit{Also, \emoji{521}\emoji{521}~for data-driven styles!}'' in an issue%\footnote{\url{https://github.com/mapbox/mapbox-gl-native/issues/8303}} 
~where the clapping hands represent a round of applause (i.e., the \textit{sentiment expression} intention) meanwhile improve the expressiveness of the statement by replacing the corresponding textual words. With a comparison, we adopt \textit{sentiment expression} as the main intention of \emoji{521} in this post.

\subsubsection{Scope of context} To determine the role of emojis in a post, it is not sufficient to study the sentence containing the emoji because the content can be loosely organized, and the emoji can be connected to a sentence far from it. For example, in the first paragraph of an issue,%\footnote{\url{https://github.com/mbilker/cypher/issues/16}} 
~i.e., ``\textit{Sorry, but I'm unfamiliar with Javascript apps. How do you install the cypher plugin? Step-by-step instructions might be useful for people like me who are not familiar with how this works and too dumb to figure this out. \emoji{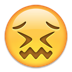}}'', the confounded face on the verge of tears is used to strengthen the sentiment expressed by the word ``sorry'' in the beginning. 

%A common practice to determine the role of an element in texts is to select the nearest sentence\reminder{cite}. However, we find it not sufficient to understand the intention of using emojis in a post on GitHub for two main reasons. First, the length of a post on GitHub is not strictly restricted and the content can be loosely organized. The meaning of an emoji can be connected to a farther sentence instead of the nearest one. For example, in the first paragraph of an issue,\footnote{\url{https://github.com/mbilker/cypher/issues/16}} i.e., ``\textit{Sorry, but I'm unfamiliar with Javascript apps. How do you install the cypher plugin? Step-by-step instructions might be useful for people like me who are not familiar with how this works and too dumb to figure this out. \emoji{852}}'', the confounded face on the verge of tears is used to strengthen the sentiment expressed by the word ``sorry'' in the beginning. 

Additionally, due to the orientation of project development and collaboration, the context of emojis on GitHub consists of not only texts but also operational behaviors.  A typical scenario is that a single emoji constitutes a post without any textual words such as the pull request comment ``\emoji{132}'',%\footnote{\url{https://github.com/dcos/dcos-ui/pull/1036#issuecomment-244138321}}
~which demonstrates sparkles. By investigating the original pull request, the previous replies, and behaviors including adding labels and merging a commit around this post, we find that the \emoji{132} represents a positive attitude to a commit and categorize its intention to \textit{sentiment expression}. %\reminder{automation of classification is challenging}

\subsubsection{Intention vs. position}
The position of the emoji in text can also affect the intention categorization. In specific, when a sentimental emoji appears in the beginning of an expression followed by sentimental texts (e.g., ``\textit{\emoji{519}~LGTM, thanks!}''%\footnote{https://github.com/purescript-contrib/purescript-freet/pull/5\#issuecomment-254262387}
), it can be labelled to either of the two sub-categories of \textit{sentimental usage}. In this work, we determine to label it as \textit{sentiment expression} for no sentiment has been conveyed before the appearance of this emoji. As a comparison, the thumbs-up gesture in ``\textit{LGTM, thanks! \emoji{519}}'' will be labelled with a \textit{sentiment strengthening} intention.
%When a sentimental emoji appears before the text, we label it to sentiment expression. \reminder{example 1}%strengthening considering its effects on the textual information.

\subsubsection{Definition vs. context-based understanding}
The embedding-based analysis in Section~\ref{sec:interpretation} has revealed domain-specific semantics of emojis determined by the technical context on GitHub. In the manual annotation process, we confirmed such usage of emojis (such as \emoji{469}, \emoji{944}, and \emoji{910}) in our sampled posts. However, we also find that such emojis can have other meanings. For example, the rocket can mean launch (``\textit{Launching soon~\emoji{910}}''%\footnote{https://github.com/waffleio/waffle.io/issues/1297\#issuecomment-259605291}
) as well as acceleration (``\textit{It uses the new osmium and its nodejs bindings for~\emoji{910}~performance.}''%\footnote{\url{https://github.com/mvexel/OSMQualityMetrics}}
), which is not domain specific on GitHub. Additionally, the emoji can be used as a symbol regardless of its meaning (e.g., ``\textit{\emoji{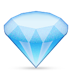}\emoji{910}~gym~has moved to the~fastlane~main~repo~\emoji{910}\emoji{584}}''%\footnote{\url{https://github.com/fastlane/gym}}
). 

Another observation comes from the emoji \emoji{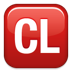}, which is known as clear button by definition and can find its application to mean ``clear''.%\footnote{\url{https://github.com/vgstation-coders/vgstation13/pull/12421}} 
~This emoji can also be interpreted to be ``change log'' or ``change list'' in the vision control system. For example,  a list of change logs are wrapped by ``\emoji{171}'' and ``/\emoji{171}'' in a pull request,%\footnote{\url{https://github.com/ParadiseSS13/Paradise/pull/5704}} 
~where we annotate its intention as \textit{statement enriching}.

Such observation indicates ambiguity of emojis on GitHub, and the meaning should be determined by both the contextual information and the intention of using the emoji. %In other words, \textcolor{blue}{the intention of using an emoji should be considered in understanding semantics and sentiments of an emoji as a supplement to text analysis.}

\subsubsection{Sentiment expression vs. atmosphere adjusting}
To distinguish the two different intentions in practice is challenging for emojis especially facial emojis. It is not rare that such emojis are used merely to make the expression more friendly instead of showing emotions or attitudes. In the pull request ``\textit{... Please take a look and check if it's worth it having it here~\emoji{834}}'',%\footnote{\url{https://github.com/datasciencebr/serenata-de-amor/pull/226}} 
~the smiling face with open mouth and smiling eyes is used to adjust the tone in order not to be serious, just like the facial expression in face-to-face communications. The intention of this \emoji{834} is categorized to \textit{atmosphere adjusting} while in the following example, the pull request ``\textit{... All covered with new tests~\emoji{834}}'',%\footnote{\url{https://github.com/beauby/jsonapi/pull/27}} 
~it is \textit{sentiment expression} because this emoji expressed happiness for adding new methods covered with new tests. In a word, the contextual information is leveraged to determine the real intention of such emojis. 

%Although sentimental emojis can be used in these two situations, emojis showing clear attitudes such as the thumbs up (\emoji{519}) are used to express or strengthen sentiments. % and the party popper (\emoji{331}) are seldom used to merely 

\subsubsection{Unusual usage}
Emojis can be used in an unusual way, which is reasonable because there are no strict rules and people are still experimenting with emojis. However, unusual usage can introduce obstacles to understand the emoji and even the post. For example, the emoji of thumbs up in the issue ``\textit{I have tried to implement the ABC algorithm, but am getting problem with \emoji{519}}'' followed by a line of code%\footnote{\url{https://github.com/prash071/Soft-Computing-Algorithms/issues/1}} 
~is challenging for us to figure out. As the other intention categories can not apply to this case, we suppose the \emoji{519} is used to referring to the following code as the mentioned problem where \emoji{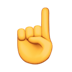} and \emoji{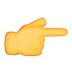} are more suitable, and to enrich the statement. 

\subsubsection{Criteria consistency}
Considering that some of the criteria emerge during the annotation process, we looked back to the posts after the process and made necessary adjustments to ensure the consistency of the categorization.

\subsection{Intention Distribution}
\begin{figure}[hbt]
\center
  \includegraphics[width=1\columnwidth]{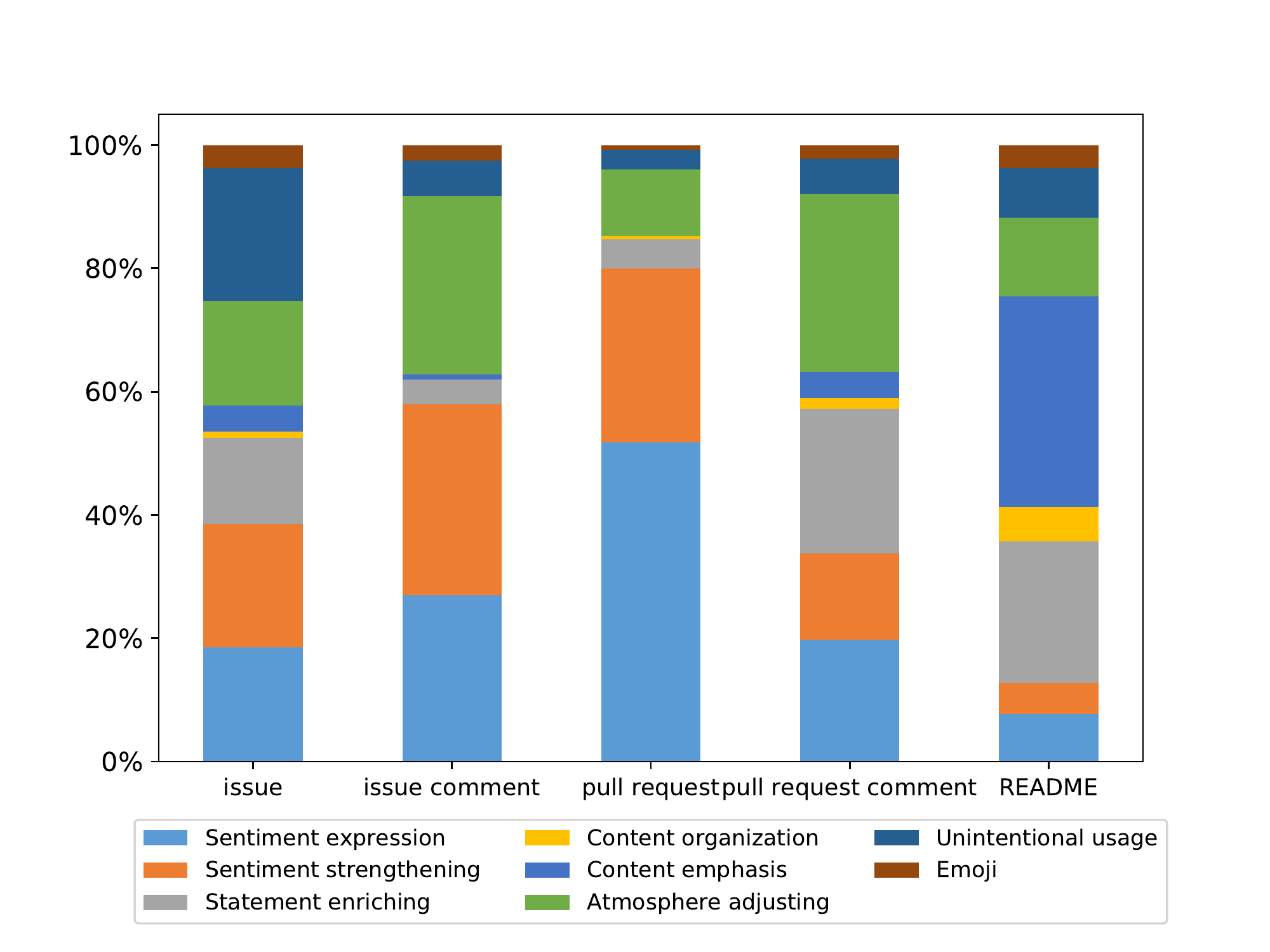}
  \caption{Distribution of intentions.}
  \label{fig:intention}
\end{figure}

The intentions of using emojis in the sampled posts are presented in Fig.~\ref{fig:intention}. We next briefly report the distribution of intentions in different posts, respectively.
%In the proposed intentions, emojis used for \textit{sentiment expression} and \textit{sentiment strengthening} occupy 44.60\% of the labelled emoji occurrences in total. For the rest, emojis are used to adjust atmosphere (19.65\%), to enrich statement (13.85\%), unintentionally (8.85\%), to help emphasize content (8.70\%), to be the symbol of emoji (2.60\%), and to help organize content (1.75\%). 
%For each of the post type, the distribution of intentions in the samples can represent the population (only English posts) with a 95\% confidence level and a 5\% confidence interval. We next discuss the distribution of intentions in different posts, respectively.

%\begin{table*}[htbp]
%\centering
%\caption{Intentions of Using Emojis.}
%\begin{tabular}{|c|c|c|c|c|c|c|c|c|}
%\hline
%\multirow{2}{*}{Post type} & \multicolumn{2}{c|}{Sentiment} & \multirow{2}{*}{Statement enriching} & \multirow{2}{*}{Organization} & \multirow{2}{*}{Emphasis} & \multirow{2}{*}{Atmosphere adjusting} & \multirow{2}{*}{Unintentional} & \multirow{2}{*}{Emoji} \\\cline{2-3}
% & Expression & Strengthening & & & & & & \\\hline
%issue	&	74	&	80	&	56	&	4	&	17	&	68	&	86	&	15 \\\hline
%pull request comment	&	207	&	113	&	19	&	2	&	0	&	43	&	13	&	3 \\\hline
%issue comment	&	108	&	124	&	16	&	0	&	3	&	116	&	23	&	10 \\\hline
%pull request	&	79	&	56	&	94	&	7	&	17	&	115	&	23	&	9 \\\hline
%README	&	31	&	20	&	92	&	22	&	137	&	51	&	32	&	15 \\\hline
%Total	&	499	&	393	&	277	&	35	&	174	&	393	&	177	&	52 \\\hline
%\end{tabular}
%\label{tab:intention}
%\end{table*}

\subsubsection{Issue} In issues, emojis are mostly used unintentionally with a 21.50\% proportion. This is because emojis often occur in pasted logs and code in raised issues, which echoes the extremely dense usage of emojis in the middle of issue posts (see Fig.~\ref{fig:post_pos}). Such finding also evidences that emojis are widely adopted in multiple scenarios in software engineering. Followed intentions are sentiment strengthening (20.00\%), sentiment expression (18.50\%), atmosphere adjusting (17.00\%), and statement enriching (14.00\%). 

The intention of representing emojis, although is the last but one intention with a 3.75\% proportion, is mostly seen in issues as well as README. Such intention often occur in emoji related contexts and in opinion consultations such as ``\textit{Click on +\emoji{833}~to add your votes}''%\footnote{https://github.com/ailin-nemui/scripts.irssi.org/issues/6\#issuecomment-242191159} 
~where \emoji{833} is one of the six reactions provided by GitHub.

\subsubsection{Pull request} The four main intentions of using emojis in pull requests are atmosphere adjusting (28.75\%), statement enriching (23.50\%), sentiment expression (19.75\%), and sentiment strengthening (14.00\%). In comparison with issues, the unintentional use is significantly reduced, and the emojis are more used to enrich statement and adjust atmosphere. On possible explanation of such difference is the relatively narrow scope of target audience (i.e., the project owner) and explicit requests (i.e., to get the contributed code pulled and merged to the repository) of pull requests.

\subsubsection{Comments} Sentimental usage of emojis is dominant in the two types of comments especially in pull request comments, where more than half of emojis (i.e., 51.75\%) are used to express sentiments and 28.25\% to strengthen sentiments. This is understandable because opinions are needed to response to raised questions, discussions, plans, and implementations. In the comments of issues, another main intention of using emojis is atmosphere adjusting (29.00\%), possibly due to the demand of smoothing communications in various discussion contexts. 

\subsubsection{README} Instead of sentimental usage, the main intention of using emojis in README is content emphasis, occupying 34.25\% of emoji occurrence. Because of the nature to communicate expectations for projects, README files are often with long texts and multiple points. Using emojis can effectively help attract attention to the emphasized points from the massive content. Another observation is that emojis contribute in hierarchical lists or checklists with a probability of 5.50\%. These two intentions of using emojis in README are significantly higher than in other posts.

\noindent \textbf{Summary}. 
Knowing why an emoji is used is of significance to understand the real meaning of the emoji and the expression with the  emoji. In this section, we proposed a taxonomy of emoji usage intention on GitHub based on our understanding and a manual annotation task conducted with 2,000 posts. The ambiguity of emojis, the diversity of emoji selection, and the complicated context of the emoji can affect the intention determination process and make it challenging. The generated taxonomy as well as the demonstrated usage examples can provide a guidance for intention recognition before leveraging emojis in further analysis. 

By completing the annotation task, we obtain the distribution of emoji usage intentions in different posts. We find that emojis are heavily used to express or strengthen sentiments in conversational posts especially in comments, while emojis are mainly used to emphasize the content in README, % as it is the first item a visitor will see when visiting a project, 
which often contains a mass of information. We also find that emojis are widely used to smooth communications to develop a positive and friendly atmosphere of the open-source collaboration platform.
%Emojis are sometimes socio-cultural norms, which should be interpreted with its context including sentence, the post, and even the whole conversation. 

%other taxonomy

%% file: sections/implication.tex
\section{Implications}\label{sec:implication}

\textbf{Supplement for sentiment analysis}. %Dictionary expansion for sentiment analysis
Understanding the thoughts and sentiments of developers has always been an active direction in software engineering research. However, sentiment analysis in SE tasks is challenging due to the unreliable results provided by existing tools~\cite{lin2018sentiment, Jongeling2017On, Novielli2015The, Naaman:2018ce}. One difficulty in automated sentiment analysis with lexical approaches such as SentiStrength~\cite{Thelwall2011Sentiment} in the SE domain is the lack of sentimental words in dictionary. In our study, emojis are found to be widely leveraged in GitHub posts to not only strengthen but also express sentiments as substitute of plain text, which implies the necessity to enlarge the dictionary of sentimental words with emojis. In addition, it is also possible to use emojis as weak sentiment labels~\cite{ZhaoDWX12, DBLFelboMSRL17} in distant-supervised learning.% automated classification is challenging.

\textbf{Status sensor for contribution activeness management}. 
The activity of contributors of open source projects can be affected by their emotion and psychological status. For example, contributors are more likely to become inactive when they express strong positive or negative emotions~\cite{Garcia2013The}. \cite{Vidal:2016fl} investigated food-related emotional experiences by analyzing emoji usage on Twitter, and find emoji usage can reflect the status of users such as being accompanied or not. Emojis can be leveraged as a sensor of contributors' status on open source platforms such as GitHub. In addition, developers can add emojis to their conversations to promote a mild and friendly atmosphere for the platform~\cite{hatfield1993emotional}.%, as has been observed in this work. %Emotional Contagion

\textbf{Visual design for problem solving and project collaboration}.
The analysis in Section~\ref{sec:dis} implies that emoji usage can be related with participation in issues and activeness of developers, possibly due to the eye-catching visual design which makes the issue more readable and less boring. Such initial findings encourage further research of the role of emojis in helping problems to be solved and promoting project collaboration in the open source community. Broadly, adding visual designs into traditionally text-heavy tasks not only adds fun to the work, but may also help engage users in the tasks and even improve the quality of work \cite{sutcliffe2009designing}. On the other side, GitHub and other developer communities like StackOverflow may consider to add more visual features to attract users into discussions, such as animations or GIF images.
%Recent interface designs have explored the use of emojis as passwords \cite{kraus2016implications}, as user names,\footnote{\url{http://www.businessinsider.com/emojli-network-where-usernames-are-emoji-2014-6}, retrieved in September 2017.} or even as part of programming languages.\footnote{\url{http://www.emojicode.org/}, retrieved in Semptember 2017.} Our work provides direct evidence of the positive effect of using emojis in user engagement and problem solving. 

In fact, the current low ratio of posts containing emojis indicates a great opportunity for the GitHub community to promote emojis in the conversation, through the designs of recommender systems or specialized interfaces. %It is also worth noting that issues with emojis did not actually attract the first comment sooner. \textcolor{red}{This is due to the lack of a mechanism for the users to find these emoji posts, and therefore implies the opportunity to design such a mechanism to quickly access posts with certain emojis (e.g., like the design of hashtags on Twitter).}

\textbf{Community-specific and identity-based design}.
Although emojis have evolved into an ubiquitous language, we have seen that a specific user group tend to use emojis and interpret emojis in different ways. %The same emojis may have quite different meanings on GitHub and Twitter. %Even programmers of different languages have different preference on emojis. 
The community-specific interpretations of emojis has become the norm of the social group and are strongly tied to the social identity of the community~\cite{gumperz1982language}. This indicates the rationale and opportunity of designing community-specific emojis, emojis that advocate for the unique identity and culture of the community. %Indeed, recently there have been emojis designed for sports teams,\footnote{\url{http://www.sportingnews.com/nfl/news/nfl-emojis-bill-belichick-richard-sherman-peyton-manning-aaron-rodgers-sportsmanias/1cp00xisf6ci114vb0qgjovqeq}, retrieved in September 2017.} corporations,\footnote{\url{https://www.fastcompany.com/3049550/check-your-head-new-mcdonalds-ad-takes-emoji-into-the-real-world}, retrieved in September 2017.} universities,\footnote{\url{https://www.ecampusnews.com/mobile-technology-2/emojis-alumni-engagement/}, retrieved in September 2017.} and even social events.\footnote{\url{http://ampthemag.com/idea-factory/steal-this-idea-customized-emojis-for-your-event/}, retrieved in September 2017.} There is a new challenge for digital design, which needs to be timely, adaptively, and perhaps democratically \cite{sutton2017provocation}. Yet we have to be careful not to bring separation to the community and sacrifice the inclusiveness.

%\textbf{Indicator of health and stress of developers}.
%Emoji usage could influence many health issues and medical aspects. \cite{Vidal:2016fl} investigated food-related emotional experiences by analyzing emoji usage on Twitter.

% working style, personality 

%\noindent $\bullet$ \textbf{/efficiency and interests}

%\noindent $\bullet$ \textbf{/Possible factors}.

%\noindent $\bullet$ \textbf{/Emojis as a supplement to text analysis}.

%A common method in these studies is to analyze texts with natural language process techniques such as in~\cite{jurado2015sentiment}. However, with the sharp increase of emoji usage in programming community like GitHub, the rich semantics and sentiments of emojis should not be neglected. Besides, emojis can be a substitute of plain text in some cases like ``\emoji{331}\emoji{331}\emoji{331}'' commenting a bug fixing commit without any text word, indicating the necessity of interpreting emojis as a supplement to text analysis to study personalities of developers and culture of the platform.

%% file: sections/threats.tex
\section{Discussion}\label{sec:threats}
\subsection{Threats to Validity}
The quality of data can be determined by the design of GitHub API and the collection strategy of GHTorrent. For example, some links can not be reached because users changed their usernames.\footnote{\url{https://help.github.com/articles/what-happens-when-i-change-my-username/}, retrieved in August 2018} The data pre-processing, during which we extract texts from the crawled markdown files, replace code, URLs, and pictures with labels, and filter duplicates and spam posts containing obviously redundant emojis, can also lead to loss of information. 

%The measure of emoji sentiments in Section~\ref{sec:senti} leverages the LIWC tool to generate sentiment scores of neighboring words of emojis. We claim that this analysis is to demonstrate the sentiment distribution of emojis on GitHub, and the specific sentiments of emojis in each post should be determined by both the contextual texts and the intention of use. 

The proposed intention taxonomy is based on our understanding and observation from the selected posts, which can be affected from three factors. First, although we select 400 posts for each post type for statistical significance, we can not guarantee the full coverage of emoji usage types. We claim the extensibility of the taxonomy for occurrence of new uses. Second, the selected posts are all in English, hence the derived distribution of emoji usage intentions can not represent that in non-English posts. Third, although the annotation task was conducted by two of the authors independently followed by a discussion to address discrepancies, mistakes may still occur because the gap between the reader and the writer can always exist. The results can be influenced by subjective opinions of the annotators and their language, education and culture backgrounds. 

The findings and the proposed taxonomy can not be directly generalized to the whole field of software engineering, due to the following two reasons. On the one hand, the studied emojis are from free communication texts on GitHub, not including other artifacts such as code and documentation. On the other hand, emojis in communications on other platform can show different characteristics, hence need to be studied with the specific contexts.

%\noindent $\bullet$ \textbf{Data quality}. Due to various reasons, some issues can not be obtained through stored url. For example, some links become invalid because users changed their usernames.\footnote{https://help.github.com/articles/what-happens-when-i-change-my-username/, retrieved in March 2018} The data collection strategy of Gotorrent

%\noindent $\bullet$ \textbf{Compatibility issues}. We use a chrome browser to open the crawled urls of posts and find several cases of rendering failure. We use the emojipedia website to search the emoji.

\subsection{Future Work}
%In this work we provide a descriptive analysis of 
The use of emojis can be affected by multiple factors, including the popularity of the project, the topic and atmosphere of the discussion, the background and personality of the user, the time of posting, etc. %Meanwhile, as implicated by the initial analysis in  
We plan to study emoji usage in depth by modeling the factors to uncover such relations. Another direction is to automate the intention categorization by developing machine learning models.

%% file: sections/conclusion.tex
\section{Conclusion}\label{sec:conclusion}

We presented a large-scale empirical study on how developers use emojis in their communications on GitHub. The data set involves free texts including conversations and README from 3.09 million GitHub projects covering 3.95 million users spanning from January 2012 to June 2017. We found that developers show domain specific preference, usage, and understanding of emojis in their communications in comparison to common Internet users. %We presented interesting usage patterns from dimensions of density, position, and selection of emojis in posts. 
We conducted a manual annotation task with 2,000 posts to propose a taxonomy of emoji usage intention for GitHub users. Results show that in addition to be widely used to smooth communications, emojis are heavily used to help express sentiments in conversations especially in comments and are mainly used as eye-catching symbols in README files. Emojis may be used as sentiment sensors of developers and should be considered in textual analysis in the software engineering field.

%We found that developers do embrace the digital fashion and use a large variety of emojis in their communications. Compared to common Internet users, the developer community presents a positive and supportive culture. We presented a statistical analysis on the effect of using emojis in programming-related issues. Results show that including emojis into the description of an issue increases the likelihood of the issue being commented and resolved. Emojis may be used as an instrument both to engage the users into collaborative tasks within the community and to attract new participants from outside the community. 

%% file: github.bbl
\begin{thebibliography}{10}
\providecommand{\url}[1]{#1}
\csname url@samestyle\endcsname
\providecommand{\newblock}{\relax}
\providecommand{\bibinfo}[2]{#2}
\providecommand{\BIBentrySTDinterwordspacing}{\spaceskip=0pt\relax}
\providecommand{\BIBentryALTinterwordstretchfactor}{4}
\providecommand{\BIBentryALTinterwordspacing}{\spaceskip=\fontdimen2\font plus
\BIBentryALTinterwordstretchfactor\fontdimen3\font minus
  \fontdimen4\font\relax}
\providecommand{\BIBforeignlanguage}[2]{{%
\expandafter\ifx\csname l@#1\endcsname\relax
\typeout{** WARNING: IEEEtran.bst: No hyphenation pattern has been}%
\typeout{** loaded for the language `#1'. Using the pattern for}%
\typeout{** the default language instead.}%
\else
\language=\csname l@#1\endcsname
\fi
#2}}
\providecommand{\BIBdecl}{\relax}
\BIBdecl

\bibitem{Vidal:2016fl}
L.~Vidal, G.~Ares, and S.~R. Jaeger, ``{Use of emoticon and emoji in tweets for
  food-related emotional expression},'' \emph{Food Quality and Preference},
  vol.~49, pp. 119--128, 2016.

\bibitem{chen2018twitter}
Y.~Chen, J.~Yuan, Q.~You, and J.~Luo, ``Twitter sentiment analysis via bi-sense
  emoji embedding and attention-based lstm,'' \emph{arXiv preprint
  arXiv:1807.07961}, 2018.

\bibitem{chen18}
Z.~Chen, X.~Lu, W.~Ai, H.~Li, Q.~Mei, and X.~Liu, ``Through a gender lens:
  Learning usage patterns of emojis from large-scale {Android} users,'' in
  \emph{Proceedings of the 27th Web Conference2016}, 2018, pp. 763--772.

\bibitem{marengo2017assessing}
D.~Marengo, F.~Giannotta, and M.~Settanni, ``Assessing personality using emoji:
  An exploratory study,'' \emph{Personality and Individual Differences}, vol.
  112, pp. 74--78, 2017.

\bibitem{Lu:UbiComp16}
X.~Lu, W.~Ai, X.~Liu, Q.~Li, N.~Wang, G.~Huang, and Q.~Mei, ``Learning from the
  ubiquitous language: an empirical analysis of emoji usage of smartphone
  users,'' in \emph{Proceedings of 2016 {ACM} International Joint Conference on
  Pervasive and Ubiquitous Computing}, 2016, pp. 770--780.

\bibitem{siadati2017modelling}
R.~Siadati, P.~Wernick, and V.~Veneziano, ``Modelling politics in requirements
  engineering: Adding emoji to existing notations,'' \emph{arXiv preprint
  arXiv:1703.06101}, 2017.

\bibitem{Islam:2018hz}
M.~R. Islam and M.~F. Zibran, ``{Leveraging automated sentiment analysis in
  software engineering},'' in \emph{Proceedings of the 14th International
  Conference on Mining Software Repositories}, 2017, pp. 203--214.

\bibitem{Novielli2015The}
N.~Novielli, F.~Calefato, and F.~Lanubile, ``The challenges of sentiment
  detection in the social programmer ecosystem,'' in \emph{International
  Workshop on Social Software Engineering}, 2015, pp. 33--40.

\bibitem{lin2018sentiment}
B.~Lin, F.~Zampetti, G.~Bavota, M.~Di~Penta, M.~Lanza, and R.~Oliveto,
  ``Sentiment analysis for software engineering: How far can we go?'' in
  \emph{Proceedings of 40th International Conference on Software Engineering},
  2018, pp. 94--104.

\bibitem{LjubesicF16}
N.~Ljubesic and D.~Fiser, ``A global analysis of emoji usage,'' in
  \emph{Proceedings of the 10th Web as Corpus Workshop}, 2016, pp. 82--89.

\bibitem{Ai:2017wx}
W.~Ai, X.~Lu, X.~Liu, N.~Wang, G.~Huang, and Q.~Mei, ``{Untangling emoji
  popularity through semantic embeddings},'' \emph{Proceedings of the 11th
  International AAAI Conference on Web and Social Media}, pp. 2--11, 2017.

\bibitem{al2015forms}
F.~Al~Rashdi, ``Forms and functions of emojis in whatsapp interaction among
  omanis,'' 2015.

\bibitem{zhou2017goodbye}
R.~Zhou, J.~Hentschel, and N.~Kumar, ``Goodbye text, hello emoji: Mobile
  communication on {WeChat} in china,'' in \emph{Proceedings of the 2017 CHI
  Conference on Human Factors in Computing Systems}, 2017, pp. 748--759.

\bibitem{barbieri2016does}
F.~Barbieri, F.~Ronzano, and H.~Saggion, ``What does this emoji mean? a vector
  space skip-gram model for twitter emojis.'' in \emph{Language Resources and
  Evaluation Conference}, 2016.

\bibitem{Novak:2015}
P.~K. Novak, J.~Smailovic, B.~Sluban, and I.~Mozetic, ``Sentiment of emojis,''
  \emph{PloS One}, vol.~10, no.~12, 2015.

\bibitem{Hu2017Spice}
T.~Hu, H.~Guo, H.~Sun, T.~T. Nguyen, and J.~Luo, ``Spice up your chat: The
  intentions and sentiment effects of using emoji,'' in \emph{Proceedings of
  the 11th International AAAI Conference on Web and Social Media}, 2017, pp.
  101--111.

\bibitem{Cramer:2016}
H.~Cramer, P.~de~Juan, and J.~R. Tetreault, ``Sender-intended functions of
  emojis in {US} messaging,'' in \emph{Proceedings of the 18th International
  Conference on Human-Computer Interaction with Mobile Devices and Services},
  2016, pp. 504--509.

\bibitem{PohlDR17}
H.~Pohl, C.~Domin, and M.~Rohs, ``Beyond just text: semantic emoji similarity
  modeling to support expressive communication \emoji{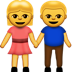} \emoji{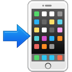}
  \emoji{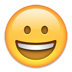},'' \emph{ACM Transactions on Computer-Human Interaction},
  vol.~24, no.~1, pp. 6:1--6:42, 2017.

\bibitem{sinha2016analyzing}
V.~Sinha, A.~Lazar, and B.~Sharif, ``Analyzing developer sentiment in commit
  logs,'' in \emph{Proceedings of the 13th International Conference on Mining
  Software Repositories}, 2016, pp. 520--523.

\bibitem{robinson2016developer}
W.~N. Robinson, T.~Deng, and Z.~Qi, ``Developer behavior and sentiment from
  data mining open source repositories,'' in \emph{System Sciences (HICSS),
  2016 49th Hawaii International Conference on}, 2016, pp. 3729--3738.

\bibitem{pletea2014security}
D.~Pletea, B.~Vasilescu, and A.~Serebrenik, ``Security and emotion: sentiment
  analysis of security discussions on github,'' in \emph{Proceedings of the
  11th working conference on mining software repositories}, 2014, pp. 348--351.

\bibitem{JongelingDS15}
R.~Jongeling, S.~Datta, and A.~Serebrenik, ``Choosing your weapons: On
  sentiment analysis tools for software engineering research,'' in \emph{2015
  {IEEE} International Conference on Software Maintenance and Evolution}, 2015,
  pp. 531--535.

\bibitem{AhmedBIR17}
T.~Ahmed, A.~Bosu, A.~Iqbal, and S.~Rahimi, ``Senticr: a customized sentiment
  analysis tool for code review interactions,'' in \emph{Proceedings of the
  32nd {IEEE/ACM} International Conference on Automated Software Engineering},
  2017, pp. 106--111.

\bibitem{DBLP:conf/esem/IslamZ17}
M.~R. Islam and M.~F. Zibran, ``A comparison of dictionary building methods for
  sentiment analysis in software engineering text,'' in \emph{2017 {ACM/IEEE}
  International Symposium on Empirical Software Engineering and Measurement},
  2017, pp. 478--479.

\bibitem{DBLP:conf/msr/GuzmanAL14}
E.~Guzman, D.~Az{\'{o}}car, and Y.~Li, ``Sentiment analysis of commit comments
  in github: an empirical study,'' in \emph{11th Working Conference on Mining
  Software Repositories}, 2014, pp. 352--355.

\bibitem{rousinopoulos2014sentiment}
A.-I. Rousinopoulos, G.~Robles, and J.~M. Gonz{\'a}lez-Barahona, ``Sentiment
  analysis of free/open source developers: preliminary findings from a case
  study,'' \emph{Revista Electronica de Sistemas de Informa{\c{c}}ao}, vol.~13,
  no.~2, p.~1, 2014.

\bibitem{ZhaoDWX12}
J.~Zhao, L.~Dong, J.~Wu, and K.~Xu, ``{MoodLens: an emoticon-based sentiment
  analysis system for Chinese Tweets},'' in \emph{The 18th {ACM} {SIGKDD}
  International Conference on Knowledge Discovery and Data Mining}, 2012, pp.
  1528--1531.

\bibitem{DBLFelboMSRL17}
B.~Felbo, A.~Mislove, A.~S{\o}gaard, I.~Rahwan, and S.~Lehmann, ``Using
  millions of emoji occurrences to learn any-domain representations for
  detecting sentiment, emotion and sarcasm,'' in \emph{Proceedings of the 2017
  Conference on Empirical Methods in Natural Language Processing}, 2017, pp.
  1615--1625.

\bibitem{Gousi13}
G.~Gousios, ``The ghtorrent dataset and tool suite,'' in \emph{Proceedings of
  the 10th Working Conference on Mining Software Repositories}, 2013, pp.
  233--236.

\bibitem{githubapi}
\BIBentryALTinterwordspacing
``{REST API v3}.'' [Online]. Available: \url{https://developer.github.com/v3/}
\BIBentrySTDinterwordspacing

\bibitem{pavalanathan2015emoticons}
U.~Pavalanathan and J.~Eisenstein, ``Emoticons vs. emojis on twitter: A causal
  inference approach,'' \emph{arXiv preprint arXiv:1510.08480}, 2015.

\bibitem{wilcoxon1945individual}
F.~Wilcoxon, ``Individual comparisons by ranking methods,'' \emph{Biometrics
  bulletin}, vol.~1, no.~6, pp. 80--83, 1945.

\bibitem{time2014emoji}
\BIBentryALTinterwordspacing
K.~Steinmetz, ``{TIME Exclusive: Here Are Rules of Using Emoji You Didn't Know
  You Were Following}.'' [Online]. Available:
  \url{http://time.com/2993508/emoji-rules-tweets/}
\BIBentrySTDinterwordspacing

\bibitem{Naaman:2018ce}
N.~Na'aman, H.~Provenza, and O.~Montoya, ``{Varying Linguistic Purposes of
  Emoji in (Twitter) Context},'' in \emph{Proceedings of ACL 2017, Student
  Research Workshop}, 2018, pp. 136--141.

\bibitem{jackson2016pragmatics}
R.~C. Jackson \emph{et~al.}, ``The pragmatics of repetition, emphasis and
  intensification,'' Ph.D. dissertation, Salford, 2016.

\bibitem{mikolov2013distributed}
T.~Mikolov, I.~Sutskever, K.~Chen, G.~S. Corrado, and J.~Dean, ``Distributed
  representations of words and phrases and their compositionality,'' in
  \emph{Advances in neural information processing systems}, 2013, pp.
  3111--3119.

\bibitem{Jongeling2017On}
R.~Jongeling, P.~Sarkar, S.~Datta, and A.~Serebrenik, ``On negative results
  when using sentiment analysis tools for software engineering research,''
  \emph{Empirical Software Engineering}, vol.~22, no.~5, pp. 2543--2584, 2017.

\bibitem{Thelwall2011Sentiment}
M.~Thelwall, K.~Buckley, and G.~Paltoglou, ``Sentiment strength detection for
  the social web,'' \emph{Journal of the Association for Information Science \&
  Technology}, vol.~63, no.~1, pp. 163--173, 2011.

\bibitem{Garcia2013The}
D.~Garcia, M.~S. Zanetti, and F.~Schweitzer, ``The role of emotions in
  contributors activity: A case study on the gentoo community,'' in \emph{Third
  International Conference on Cloud and Green Computing}, 2013, pp. 410--417.

\bibitem{hatfield1993emotional}
E.~Hatfield, J.~T. Cacioppo, and R.~L. Rapson, ``Emotional contagion,''
  \emph{Current directions in psychological science}, vol.~2, no.~3, pp.
  96--100, 1993.

\bibitem{sutcliffe2009designing}
A.~Sutcliffe, ``Designing for user engagement: Aesthetic and attractive user
  interfaces,'' \emph{Synthesis lectures on human-centered informatics},
  vol.~2, no.~1, pp. 1--55, 2009.

\bibitem{gumperz1982language}
J.~J. Gumperz, \emph{Language and social identity}.\hskip 1em plus 0.5em minus
  0.4em\relax Cambridge University Press, 1982, vol.~2.

\end{thebibliography}
